\documentclass[aps,pra,twocolumn,showpacs,preprintnumbers,amsmath,amssymb]{revtex4-1}
\usepackage{graphicx}
\usepackage[usenames,dvipsnames]{xcolor}
\usepackage{cleveref}
\usepackage{makecell}
\usepackage{float}
\usepackage{lineno}
\usepackage[T1]{fontenc}
\usepackage[caption=false]{subfig}
\newcommand{\Tr}{\textrm{Tr}}

\newcommand{\bra}[1]{\ensuremath{\langle #1 |}}
\newcommand{\ket}[1]{\ensuremath{| #1 \rangle}}
\newcommand{\braket}[2]{\ensuremath{\langle #1 | #2 \rangle}}

\usepackage{amssymb,amsmath} 
\usepackage{amsfonts}
\usepackage{color}
\usepackage[utf8]{inputenc}
\usepackage{lmodern}
\usepackage{epsfig}
\usepackage{ulem}
\usepackage{blindtext}

\begin{document}
\title{Fidelity and entanglement entropy for infinite-order phase transitions}

\author{Jin Zhang}
\email{jin-zhang@uiowa.edu}
\affiliation{Department of Physics and Astronomy, University of Iowa, Iowa City, IA 52242, USA}
\definecolor{burnt}{cmyk}{0.2,0.8,1,0}
\def\lt{\lambda ^t}
\def\note{note}
\def\beq{\begin{equation}}
\def\enq{\end{equation}}

\date{\today}
\begin{abstract}
We study the fidelity and the entanglement entropy for the ground states of quantum systems that have infinite-order quantum phase transitions. In particular, we consider the quantum O(2) model with a spin-$S$ truncation, where there is an infinite-order Gaussian (IOG) transition for $S = 1$ and there are Berezinskii-Kosterlitz-Thouless (BKT) transitions for $S \ge 2$. We show that the height of the peak in the fidelity susceptibility ($\chi_F$) converges to a finite thermodynamic value as a power law of $1/L$ for the IOG transition and as $1/\ln(L)$ for BKT transitions. The peak position of $\chi_F$ resides inside the gapped phase for both the IOG transition and BKT transitions. On the other hand, the derivative of the block entanglement entropy with respect to the coupling constant ($S^{\prime}_{vN}$) has a peak height that diverges as $\ln^{2}(L)$ [$\ln^{3}(L)$] for $S = 1$ ($S \ge 2$) and can be used to locate both kinds of transitions accurately. We include higher-order corrections for finite-size scalings and crosscheck the results with the value of the central charge $c = 1$. The crossing point of $\chi_F$ between different system sizes is at the IOG point for $S = 1$ but is inside the gapped phase for $S \ge 2$, while those of $S^{\prime}_{vN}$ are at the phase-transition points for all $S$ truncations. Our work elaborates how to use the finite-size scaling of $\chi_F$ or $S^{\prime}_{vN}$ to detect infinite-order quantum phase transitions and discusses the efficiency and accuracy of the two methods.
\end{abstract}


\maketitle

\section{Introduction}\label{sec:introduction}

One of the main interest in condensed matter physics is to understand quantum phase transitions (QPTs) in many-body systems. Analogous to classical phase transitions, QPTs can be classified by the singularities of derivatives of the ground-state energy density: the $k$-th order QPT is signaled by a divergence or discontinuity in the $k$-th derivative of the ground-state energy density. By measuring quantities associated with these derivatives, the first-order QPT and the second-order QPT can be easily detected in experiments, as there are well developed techniques to measure local order parameters and their susceptibilities, which are associated with the first and second derivatives of the ground-state energy density, respectively. From the point of view of numerics, the ground state energy and the local observables of systems in low dimensions can also be calculated accurately by tensor-network algorithms. However, higher-order QPTs are difficult to detect using this method, as higher-order derivatives of the ground-state energy density are hard to probe in both experiments and computer programs. For infinite-order QPTs (IOQPTs), this method will not give us meaningful information.

Lots of concepts from quantum information theory have been implemented in condensed matter physics. Among them, the ground-state fidelity \cite{PhysRevE.74.031123, PhysRevE.76.061108, PhysRevA.75.032109, PhysRevA.77.032111, PhysRevA.77.012311, PhysRevA.77.062321, Dai_2010, Wang_2011(1), Wang_2011(2), PhysRevE.82.061127, PhysRevB.84.224435,  PhysRevA.89.033625, PhysRevA.97.013845, PhysRevB.95.085102, PhysRevA.98.023607, PhysRevE.98.022106, PhysRevB.100.094428} and the ground-state entanglement \cite{Osterloh2002, PhysRevA.66.032110, PhysRevLett.93.086402, PhysRevLett.93.250404, PhysRevLett.95.196406, PhysRevA.73.042320, PhysRevLett.96.116401,  PhysRevA.73.010303, PhysRevA.74.052335, PhysRevA.81.032334, PhysRevB.84.195108, LI20161066, PhysRevLett.117.206801, PhysRevLett.119.225301, PhysRevB.99.195445} have been proved to be successful to detect QPTs in various models. The fidelity method is based on the simple idea that the structure of the ground-state wavefunctions on two sides of the critical point are very different, thus there exists a drastic drop in fidelity around the critical point. This drastic drop can be characterized by a divergent quantity, the fidelity susceptibility ($\chi_F$) \cite{PhysRevE.76.022101}. One can show that $\chi_F$ has poles of one order higher than the second derivative of the ground-state energy density \cite{PhysRevE.76.022101, PhysRevA.77.032111}, thus fidelity methods work well for detecting QPTs of order less than four \cite{PhysRevE.74.031123, PhysRevA.75.032109, PhysRevA.77.032111, PhysRevA.77.012311, PhysRevA.77.062321, PhysRevA.97.013845, PhysRevA.98.023607, PhysRevE.98.022106}. Critical exponents can be extracted by finite-size scalings (FSSs) of peak heights and peak positions, which can be used to determine the order of QPTs \cite{PhysRevA.77.012311, PhysRevA.77.062321}. However, $\chi_F$ does not diverge for QPTs of order higher than three, especially for the IOQPTs. Although one can detect the IOQPTs in the $J_1$-$J_2$ Heisenberg chain using fidelity for the first-excited state \cite{PhysRevE.76.061108} or a more general definition of fidelity \cite{PhysRevB.84.224435}, the methods are specific to this model and cannot be easily generalized to other models. Making use of a pseudo spontaneous symmetry breaking in infinite matrix product states with finite bond dimension also works for IOQPTs \cite{Wang_2011(1), Wang_2011(2)}. Here we are interested in the methods based on FSS, which can be applied to experimental realizations of analog quantum simulations. Reference \cite{PhysRevB.91.014418} shows that the scaling of the peak height of $\chi_F$ does signal a BKT transition, and one can extrapolate a value close to the BKT transition point using the standard FSS of the peak position. But it is suspect that a non-divergent peak is located at the BKT transition point, as it has been shown that the finite peak of the specific heat is away from the BKT transition point and inside the gapped phase \cite{PhysRevB.42.6827, PhysRevE.101.060105}. In this paper, we take the truncated quantum O(2) model and resolve this issue based on accurate density-matrix-renormalization-group (DMRG) calculations.

As the entanglement entropy is a byproduct in DMRG calculations, it is natural to compare the fidelity methods to the entanglement methods. In the seminal work Ref. \cite{Osterloh2002}, it was shown that the singularity of the derivative of the two-site entanglement is located at the critical point of the second-order QPTs. In later works, local measures of entanglement such as single-site \cite{PhysRevA.66.032110,PhysRevLett.93.086402,PhysRevLett.95.196406,PhysRevA.73.042320} and two-site \cite{PhysRevA.66.032110,PhysRevLett.93.250404,PhysRevLett.96.116401,LI20161066} entanglement entropy were proposed for the study of finite-order QPTs. In fact, the ground-state expectation values of local observables (the ground-state energy is the expectation value of a sum of local operators for systems with short-range interactions) are linear functions of the matrix elements of few-body reduced density matrices residing at the same subsystem as the local observables \cite{PhysRevLett.93.250404}, so any local measures of entanglement that depends on the reduced density matrix should have singularities with critical exponents for finite-order QPTs \cite{PhysRevA.73.010303}. But again, the local entanglement does not have singularities in any of its finite-order derivatives for IOQPTs so it cannot be used to detect them. The successful example using single-site entanglement entropy for the one-dimensional Hubbard model is ascribed to the coincidence that the equipartition of local states is reached at the IOQPT point \cite{PhysRevA.73.042320}, and the one using two-site entanglement entropy for the $J_1$-$J_2$ Heisenberg chain is due to the coincidence that the two-site entropy can characterize the dimerized order in the gapped phase \cite{LI20161066}. The local maximum in the block entanglement entropy of the spin-$1/2$ $XXZ$ chain \cite{PhysRevA.81.032334} is found to be at the BKT point, but this is not a universal feature for BKT transitions. The local maxima in the estimated values of the central charge \cite{PhysRevB.84.195108, PhysRevB.95.085102, PhysRevB.99.195445} are also observed to be at IOQPTs, but they cannot differentiate between different types of IOQPTs. We are interested in a universal entanglement method for probing IOQPTs and extracting their critical properties. Notice that there exists a universal scaling law for the block entanglement entropy in one-dimensional quantum systems: the block entanglement entropy at a critical point diverges logarithmically with the size of the block. For gapped-to-gapped phase transitions, the phase transition point is singled out by this divergent behavior, and critical exponents can be extracted by analysis of parity-oscillation corrections \cite{PhysRevLett.104.095701, Fagotti_2011, PhysRevB.83.214425} and the FSS of peak positions. For IOQPTs from a gapped phase to a gapless phase, the block entanglement entropy may keep increasing and saturate, with no clear signals for the phase transition point. But intuitively, the peak of the derivative of the block entanglement entropy ($S^{\prime}_{vN}$) should diverge and reside at the IOQPT point. We provide a detailed analysis for the FSS of $S^{\prime}_{vN}$ in this paper.

Our goal in this work is to investigate the application of $\chi_F$ and $S^{\prime}_{vN}$ to detect and differentiate between different types of IOQPTs, and accurately locate the phase-transition points. The quantum O(2) model with spin-$1$ truncation has an IOG transition from a gapped phase into a BKT critical line, where the phase-transition point is a multicritical point connecting a Gaussian critical line and two BKT critical lines \cite{PhysRevB.67.104401,PhysRevB.103.245137}. For larger spin truncations, the model has BKT transitions. The SU(2) symmetric models such as the one-dimensional Hubbard model \cite{ovchinnikov1970excitation,PhysRevB.61.16377} and the $J_1$-$J_2$ Heisenberg chain \cite{PhysRevB.25.4925} also have this IOG transitions. The magnetic and correlation-length critical exponents for the IOG transition are the same as BKT transitions. A number of previous works assert that this IOG transition belongs to BKT-type, but the essential singularity in the correlation length at the IOG point is different from that at the BKT point. Level spectroscopy (LS) can differentiate between the two transitions and locate the phase-transition points accurately \cite{PhysRevB.103.245137}, but it needs prior knowledge about the critical properties of the model. Here we show that the FSS of the peak heights and the peak positions of $\chi_F$ and $S^{\prime}_{vN}$ can differentiate between IOG transitions and BKT transitions, and the entanglement method can locate the values of IOQPT points more accurately than the fidelity method.

The paper is organized as follows. Section~\ref{subsec:qo2model} introduces the quantum O(2) model and its phase transitions. The definition of $\chi_F$ and its relation to QPTs are described in Sections~\ref{subsec:modelgsfs}. Section~\ref{subsec:modelgsee} discusses the FSS of the peak position and the divergent behavior of the peak height for $S^{\prime}_{vN}$. Section~\ref{subsec:convergedmrg} analyzes the convergence of DMRG calculations of $\chi_F$ and $S^{\prime}_{vN}$. We discuss the numerical results in Section~\ref{sec:results}. We first give general remarks in Section~\ref{subsec:genremark}. In Sec. \ref{subsec:fss}, we show that the differences between IOG and BKT transitions are exhibited in the FSS of the peak positions and the peak heights of $\chi_F$ and $S^{\prime}_{vN}$. In the thermodynamic limit, the peak positions of $\chi_F$ are not located at IOP or BKT points, while those for $S^{\prime}_{vN}$ are. We present the FSS of crossing points of $\chi_F$ and $S^{\prime}_{vN}$ to further support our conclusions. Finally, in Section~\ref{sec:conclusion}, we summarize the main conclusions of our work.

\section{Model and Methods}\label{sec:model}

\subsection{Quantum O(2) model} \label{subsec:qo2model}
The two-dimensional classical O(2) model can be defined on a Euclidean-spacetime lattice. In the dual representation and in the time-continuum limit \cite{PhysRevA.90.063603, PhysRevD.92.076003, PhysRevD.98.094511, PhysRevB.103.245137}, the Hamiltonian formulation, or the quantum O(2) model in ($1+1$) dimensions is obtained:
\begin{align}
	\label{eq:any-spin-ham-u}
	\hat{H}_U = D \sum_{l=1}^{L} (\hat{S}_{l}^z)^2 - J \sum_{l = 1}^{L-1} \left(\hat{U}_{l}^+ \hat{U}_{l+1}^- + \hat{U}_{l}^- \hat{U}_{l+1}^+\right),
\end{align}
where $D$ and $J$ are coupling constants, $L$ is the total number of sites. $\hat{S}^z$ is an operator with its eigenvalues and eigenstates satisfying $\hat{S}^z \ket{n} = n \ket{n}$ ($n = 0, \pm 1, \pm 2, ...$), and $\hat{U}^{\pm} = \exp(\pm i \hat{\theta})$ are raising and lowering operators, $\hat{U}^{\pm} \ket{n} = \ket{n \pm 1}$. Open boundary conditions (OBCs) are considered here. We set $J = 1$ as the energy scale for all the following calculations. Without a truncation, the value of $n$ can be infinitely large. With a truncation $|n|_{max} = S$, $\hat{S}^z$ becomes the $z$-component of the spin-$S$ operator, and $\hat{U}^{\pm} \ket{\pm S} = 0$. We also consider the model with raising and lowering operators replaced by spin ladder operators $\hat{S}^{\pm}/\sqrt{S(S+1)}$,
\begin{align}
	\label{eq:any-spin-ham-ladder}
	\hat{H}_S = D \sum_{l=1}^{L} (\hat{S}_{l}^z)^2 - \frac{J}{S(S+1)} \sum_{l = 1}^{L-1} \left(\hat{S}_{l}^+ \hat{S}_{l+1}^- + \hat{S}_{l}^- \hat{S}_{l+1}^+\right).
\end{align}
For $S = 1$ or in the large-$S$ limit, $\hat{U}^{\pm} = \hat{S}^{\pm} / \sqrt{S(S+1)}$ and the two Hamiltonians are the same \cite{PhysRevB.103.245137}. $\hat{U}^{\pm}$ and $\hat{S}^z$ have the following commutation relations
\begin{eqnarray}
\left[\hat{U}^{+}, \hat{U}^{-}\right] &=& \hat{\mathcal{D}}, \label{eq:upumcommute} \\ \left[\hat{S}^z, \hat{U}^{\pm}\right] &=& \pm \hat{U}^{\pm} \label{eq:szupmcommute},
\end{eqnarray}
where $\hat{\mathcal{D}}$ only has nonzero matrix elements at the most upper-left corner, $\bra{2S+1}\hat{\mathcal{D}}\ket{2S+1} = 1$, and the most lower-right corner, $\bra{-2S-1}\hat{\mathcal{D}}\ket{-2S-1} = -1$. Equation~\eqref{eq:szupmcommute} is the same as $[\hat{S}^z, \hat{S}^{\pm}] = \pm \hat{S}^{\pm}$ for spin operators, while Eq.~\eqref{eq:upumcommute} is not the same as the commutation relation between spin ladder operators, $[\hat{S}^{+}, \hat{S}^{-}] = 2\hat{S}^z$ except for $S = 1$. Both Hamiltonian \eqref{eq:any-spin-ham-u} and Hamiltonian \eqref{eq:any-spin-ham-ladder} have an explicit global U(1) symmetry, so the total magnetization is a conserved quantum number for any spin truncation.

There is an IOG transition from the gapped large-$D$ phase into a gapless BKT critical line as we decrease $D$ for our Hamiltonian with $S = 1$, while there are BKT transitions for both Hamiltonians with $S \ge 2$ \cite{PhysRevB.103.245137}. Approaching the phase transition point from the gapped side ($D \rightarrow D^{+}_c$), the correlation length diverges in the following form:
\begin{eqnarray}
\label{eq:corrlens12}
\xi \sim (\Delta E)^{-1} \sim \begin{cases}
     \left(D-D_c\right)^{-1/2} e^{b_1/(D-D_c)}, & S = 1\\
    e^{b_S/\sqrt{D-D_c}}, & S \ge 2
\end{cases},
\end{eqnarray}
where $b_{S}$ is a non-universal constant that depends on the details of the model. Due to these essential singularities, ordinary methods of finding the phase-transition points by judging where the energy gap closes are not accurate, and finite-size effects are strong and decrease slowly due to the logarithmic scaling.

\subsection{Ground-state fidelity susceptibility} \label{subsec:modelgsfs}
The ground-state fidelity \cite{PhysRevE.74.031123, PhysRevLett.99.095701, PhysRevE.76.022101, doi:10.1142/S0217979210056335} between two ground states for coupling constants $D$ and $D+\delta$ is defined as
\begin{eqnarray}
\label{eq:gsfid}
F(D, D+\delta) = \braket{\Psi_0(D)}{\Psi_0(D+\delta)},
\end{eqnarray}
where $\ket{\Psi_0}$ is the ground state. Near the phase transition point, a small increment in $D$ can drive the system from one phase to another. If the phase transition is associated with a symmetry breaking, the structure of the ground-state wavefunction changes drastically, then the fidelity has a large drop around the phase transition. This drastic change in fidelity is characterized by a peak in the fidelity susceptibility
\begin{eqnarray}
\label{eq:fidsus}
\chi_F(L) = \lim_{\delta \rightarrow 0}\frac{-2 \ln\left(\braket{\Psi_0(D)}{\Psi_0(D+\delta)}\right)}{L \delta^2}.
\end{eqnarray}
The scaling analysis suggests that $\chi_F(L) \propto L^{1+2z-2\Delta}$ \cite{PhysRevLett.99.095701}, where $z$ and $\Delta$ are the dynamical exponent and the scaling dimension of the perturbation term (the $D$ term for our model), respectively. For BKT transitions, $z = 1, \Delta = 2$, the scaling analysis gives $\chi_F(L) \propto 1/L$. A more precise analysis based on non-Abelian bosonization concludes that the leading behavior contributed by the marginal operator at the BKT point is $\chi_F(L) \propto 1/\ln(L)$ \cite{PhysRevB.91.014418}, which can be used to detect the existence of BKT transitions. So the height of $\chi_F$ is finite in the thermodynamic limit. In fact, for IOQPTs such as BKT transitions without symmetry breaking, the structure of the ground-state wavefunction changes smoothly across the phase transition point, thus $\chi_F$ should not diverge. Using perturbation theory, one can obtain
\begin{eqnarray}
\label{eq:fsde0}
\chi_F = \frac{1}{L} \sum_{n \neq 0} \frac{|\bra{\Psi_n} \hat{H}_D \ket{\Psi_0}|^2}{(E_n-E_0)^2} = -\frac{1}{2L}\frac{\partial}{\partial E_0} \frac{\partial^2 E_0}{\partial D^2},
\end{eqnarray}
where $\ket{\Psi_n}$ is the eigenstate of $\hat{H}_{U(S)}$ with eigenenergy $E_n$, and $\hat{H}_D = \sum_l (\hat{S}^z_l)^2$. For second-order QPTs, the second derivative of the ground-state energy has poles of order one at the phase transition point where the energy gap closes, and $\chi_F$ has the same poles of order two. So $\chi_F$ is more singular than $\partial^2 E_0 / \partial D^2$ and is likely to diverge for third-order QPTs. However, for QPTs of order larger than three, $\chi_F$ is not guaranteed to be infinite. References \cite{PhysRevA.77.062321, You_2015} showed that $\chi_F$ indeed diverges for second- and third-order QPTs, but is finite for fourth- and fifth-order QPTs.

References \cite{PhysRevB.91.014418, PhysRevB.100.094428} assert that the peak position goes to the BKT point as
\begin{eqnarray}
\label{eq:fspeakfss}
D_p(L) - D_c \approx \frac{A}{\ln^2(BL)},
\end{eqnarray}
where $A$ and $B$ are constants. We should be cautious, however, when discussing the FSS of a non-divergent peak. For example, the specific heat, which is the second derivative of the free energy and is divergent for second-order phase transitions, is finite for BKT transitions and the peak position is inside the gapped phase \cite{PhysRevB.42.6827, PhysRevE.101.060105}. On one hand, the dominant scaling of $\chi_F \propto 1/\ln(L)$ comes from the marginal operator at the BKT point, so the FSS in Eq.~\eqref{eq:fspeakfss} deduced from the correlation length may also dominate the scaling of the peak position of $\chi_F$. Reference \cite{PhysRevB.100.094428} has employed Eq.~\eqref{eq:fspeakfss} to extrapolate accurate values of BKT points for clock models. On the other hand, $\chi_F$ being finite indicates that one can formulate a scaling hypothesis for the log fidelity as a function of correlation lengths $\xi(D), \xi(D+\delta)$, and show that the peak of $\chi_F$ for BKT transitions is shifted into the gapped phase \cite{PhysRevB.100.081108}, which has been checked numerically. The only possibility to resolve the contradiction is that the FSS in Eq.~\eqref{eq:fspeakfss} is valid only for intermediate system sizes. This is true because in the large $L$ limit, the finite-size effects from the marginal operator vanish and $\chi_F$ converges so that the scaling hypothesis becomes valid and determines the location of the peak. For the IOG transition, there is no reason for the peak to be around the phase-transition point. We will show that the peak position of $\chi_F$ is indeed inside the gapped phase for our model, but the FSS in Eq.~\eqref{eq:fspeakfss} can predict approximate values of the BKT points using data for intermediate system sizes.

\subsection{Derivative of the block entanglement entropy} \label{subsec:modelgsee}

By splitting the system into two parts, $\mathcal{A}$ and $\mathcal{B}$, the entanglement entropy of the ground state is
\begin{eqnarray}
S_{vN} = - \Tr[ \hat{\rho}_{\mathcal{A}} \ln(\hat{\rho}_{\mathcal{A}})],
\end{eqnarray}
where $\hat{\rho}_{\mathcal{A}} = \Tr_{\mathcal{B}} \braket{\Psi_0}{\Psi_0}$ is the reduced density matrix for block $\mathcal{A}$. We focus on the case where the system is cut in the middle and $\mathcal{A}$ is half of the system. According to the area laws of the entanglement entropy \cite{RevModPhys.82.277}, for one-dimensional quantum systems, $S_{vN}$ is finite in the gapped phase. At the critical point, the entanglement entropy diverges logarithmically with the size of the system due to conformal anomaly, which is \cite{HOLZHEY1994443, PhysRevLett.90.227902}
\begin{eqnarray}
\label{eq:eescaling}
S_{vN} = \frac{c}{6} \ln(L) + r
\end{eqnarray}
for OBCs, where $c$ is the central charge, $r$ is a non-universal constant. For IOG and BKT transitions, the entanglement entropy is finite in the gapped phase and diverges in the gapless phase, thus the phase-transition point is at the place where the derivative of $S_{vN}$ with respect to the coupling constant diverges. For finite-size systems, $S^{\prime}_{vN}$ has a peak (we take $- d S_{vN} / d D$ in this paper so $S^{\prime}_{vN}$ is positive) moving toward the phase-transition point. Based on Eq.~\eqref{eq:corrlens12} and $\xi \sim L$, one can obtain the leading behavior of the scaling of the peak position: $D_p(L) - D_c \approx b_1/\ln(L)$ for $S = 1$ and $D_p(L) - D_c \approx b_S^2/\ln^2(L)$ for $S \ge 2$. To obtain accurate results, we need to consider higher-order corrections. For $S = 1$, the leading term $b_1/\ln(L)$ is substituted back into Eq.~\eqref{eq:corrlens12} to find the main correction from $(D-D_c)^{-1/2}$ factor, and we add two more higher-order terms proportional to $1/\ln^2(L)$ and $1/\ln^3(L)$. Finally, the FSS of the peak position takes the form
\begin{eqnarray}
\label{eq:delDvslnLs1}
\nonumber D_p\left(L\right) - D_c &=& \frac{b_1}{\ln(L)} + \frac{b_1[\ln\ln(L)-\ln(b_1)]}{2\ln^2(L)+\ln(L)} \\  &&+ \frac{d_1}{\ln^2(L)} + \frac{e_1}{\ln^3(L)} + \ldots,
\end{eqnarray}
where $d_1$ and $e_1$ are constants. For $S \ge 2$, we add two correction terms proportional to $1/\ln^3(L)$ and $1/\ln^4(L)$, and the FSS is
\begin{eqnarray}
\label{eq:delDvslnLs2}
D_p\left(L\right) - D_c = \frac{b_S^2}{\ln^2(L)} + \frac{d_S}{\ln^3(L)} + \frac{e_S}{\ln^4(L)} + \ldots,
\end{eqnarray}
where $d_S$ and $e_S$ are constants depending on $S$. We will show that the extrapolated phase-transition points for $S = 1, 2, 3, 4, 5$ based on Eqs.~\eqref{eq:delDvslnLs1} and \eqref{eq:delDvslnLs2} are all close to the results from LS with differences only of order $10^{-3}$.

The peak height of $S'_{vN}$ diverges in the thermodynamic limit. To obtain the FSS, we take the derivative of both sides of Eq.~\eqref{eq:eescaling} with respect to the peak position $D_p$, then the peak height $S^{\prime *}_{vN}$ scales as
\begin{eqnarray}
\label{eq:dsdyp}
S^{\prime *}_{vN} = \frac{c}{6} \frac{1}{L} \frac{dL}{dD_p} + r^{\prime},
\end{eqnarray}
where we have assumed that $r$ is a linear function of $D_p - D_c$, which is valid as long as $|D_p - D_c|$ is small, thus $r^{\prime}$ is a constant. Combining Eqs.~\eqref{eq:delDvslnLs1} and \eqref{eq:dsdyp}, we have the FSS of $S^{\prime *}_{vN}$ for $S = 1$
\begin{eqnarray}
\label{eq:dsdys1}
S^{\prime *}_{vN} = \frac{a_1 \ln^{p_1}(L)}{1 + d'_1/\ln(L) + \ldots} + r'_1,
\end{eqnarray}
where $p_1 = 2$, and the coefficient $a_1$ is related to $b_1$ and the central charge by
\begin{eqnarray}
\label{eq:a1bcrelation}
a_1 = \frac{c}{6b_1}.
\end{eqnarray}
Combining Eqs.~\eqref{eq:delDvslnLs2} and \eqref{eq:dsdyp}, we have the result for $S \ge 2$
\begin{eqnarray}
\label{eq:dsdys2}
S^{\prime *}_{vN} = \frac{a_S \ln^{p_S}(L)}{1 + d'_S/\ln(L) + e'_S/\ln^2(L) + \ldots} +r'_S
\end{eqnarray}
where $p_S = 3$, and $a_S$ is related to $b_S$ and the central charge by
\begin{eqnarray}
\label{eq:a2bcrelation}
a_S = \frac{c}{12b_S^2}.
\end{eqnarray}
We only consider one correction term in the denominator in Eq.~\eqref{eq:dsdys1} for $S = 1$ because the curve fit is not stable with complicated higher-order corrections. Using Eqs.~\eqref{eq:dsdys1} and \eqref{eq:dsdys2} to fit the data, we can extract the values of $p_S$ to check the expected results $p_1 = 2$ for IOG transitions and $p_S = 3$ ($S \ge 2$) for BKT transitions. We also use Eqs.~\eqref{eq:a1bcrelation} and \eqref{eq:a2bcrelation} to check the values of central charge $c = 1$ after obtaining $a_S$ and $b_S$ from the curve fit. These results are only based on the renormalization-group analysis [Eq.~\eqref{eq:corrlens12}] and conformal field theory [Eq.~\eqref{eq:eescaling}], so they are universal features for IOG transitions and BKT transitions.

\subsection{Convergence of DMRG} \label{subsec:convergedmrg}
\begin{figure}
  \centering
    \includegraphics[width=0.48\textwidth]{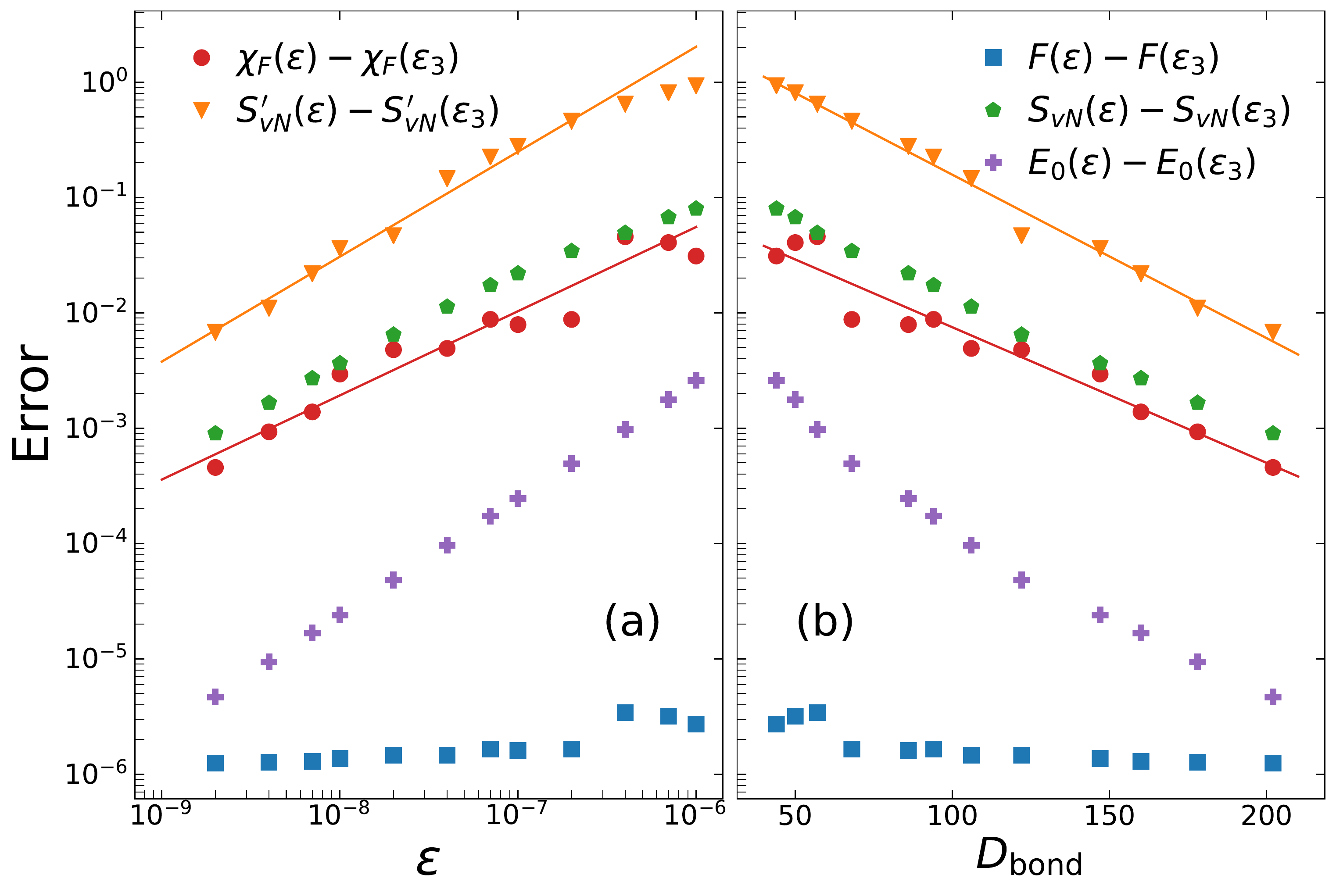}
    \caption{\label{fig:dmrgparamscaling}The dependence of the error in the Fidelity ($F$), the ground-state entanglement entropy ($S_{vN}$), the ground-state energy ($E_0$), the Fidelity susceptibility ($\chi_F$), and the derivative of $S_{vN}$ with respect to $D$ ($S_{vN}^{\prime}$) on (a) the truncation error and (b) the bond dimension in DMRG calculations. The error is obtained by subtracting the results for $\epsilon_3 = 10^{-12}$ from those for larger truncation errors. The results are for Hamiltonian \eqref{eq:any-spin-ham-u} with $S = 2, L = 384, D = 1.297$. The best linear fitting functions for log errors in $\chi_F$ and $S^{\prime}_{vN}$ are $\log_{10}|\chi_F(\epsilon) - \chi_F| = 3.1(6)+0.73(7)\log_{10}\epsilon$, $\log_{10}|S_{vN}^{\prime}(\epsilon) - S_{vN}^{\prime}| = 5.8(3)+0.91(4)\log_{10}\epsilon$, $\log_{10}|\chi_F(\epsilon) - \chi_F| = -0.95(11)-0.0118(9)D_{\rm{bond}}$, and $\log_{10}|S_{vN}^{\prime}(\epsilon) - S_{vN}^{\prime}| = 0.61(6)-0.0142(5)D_{\rm{bond}}$.
    }
  \end{figure} 
  
We perform the finite-size DMRG algorithm \cite{PhysRevLett.69.2863, PhysRevB.48.10345, SCHOLLWOCK201196} with ITensor C++ Library \cite{itensor}, which minimizes the finite-size gound-state energy by optimizing the matrix product state (MPS) \cite{PhysRevLett.75.3537} variationally. We increase the number of Schmidt states (bond dimension of MPS or $D_{\rm{bond}}$) gradually during the sweeping procedure until the truncation error is less than a preset value $\epsilon$. The number of sweeps is large enough for the difference in the entanglement entropy between the last two sweeps to be less than $10^{-11}$. We set $J = 1$ in all the calculations unless otherwise specified.

Both the calculation of $\chi_F$ and that of $S^{\prime}_{vN}$ require the determination of the ground-state wavefunctions at two close couplings $D$ and $D+\delta$. We use $\delta = 5\times 10^{-4}$ for all the calculations. The byproducts are the ground-state energy ($E_0$), the entanglement entropy ($S$), and the fidelity ($F$). In Fig.~\ref{fig:dmrgparamscaling}, we show the dependence of the errors in these quantities on the truncation error and the bond dimension of MPS for Hamiltonian \eqref{eq:any-spin-ham-u} with $S = 2$, $L = 384$, and $D = 1.297$. The errors are obtained by subtracting the results for a very small truncation error $\epsilon_3 = 10^{-12}$, which can be considered as exact, from those for truncation errors of several orders larger. We find that the error in $F$ is small and close to $10^{-6}$ for all cases considered here, and decreases slowly. The error in $E_0$ decreases as a power of $\epsilon$, consistent with the results in Refs. \cite{PhysRevB.72.180403, PhysRevB.103.245137}. The errors in $S_{vN}$, $\chi_F$, and $S^{\prime}_{vN}$ all decreases as a power of $\epsilon$ or exponentially with the bond dimension, but more than one orders larger than that in $E_0$. The value of $\chi_F$ depends on the overlap of two wavefunctions, which is more sensitive to the details of the MPS, thus has stronger fluctuations than others.

In Fig.~\ref{fig:dmrgparamscaling}(a) with both axes in the logarithmic scale, we perform linear fits for $\chi_F$ and $S^{\prime}_{vN}$ and find that $\log_{10}|\chi_F(\epsilon) - \chi_F| = 3.1(6)+0.73(7)\log_{10}\epsilon$, $\log_{10}|S_{vN}^{\prime}(\epsilon) - S_{vN}^{\prime}| = 5.8(3)+0.91(4)\log_{10}\epsilon$. Then we can estimate the errors for smaller truncation errors. The extrapolated errors in $\chi_F$ and $S^{\prime}_{vN}$ for $\epsilon_1 = 10^{-10}$ are $10^{-4.2}$ and $10^{-3.3}$, respectively. And the extrapolated errors for $\epsilon_2 = 10^{-11}$ are $10^{-4.9}$ and $10^{-4.2}$, respectively. We perform the same procedure in Fig.~\ref{fig:dmrgparamscaling}(b), and the extrapolated errors are more than one orders smaller. 

\begin{figure}
  \centering
    \includegraphics[width=0.48\textwidth]{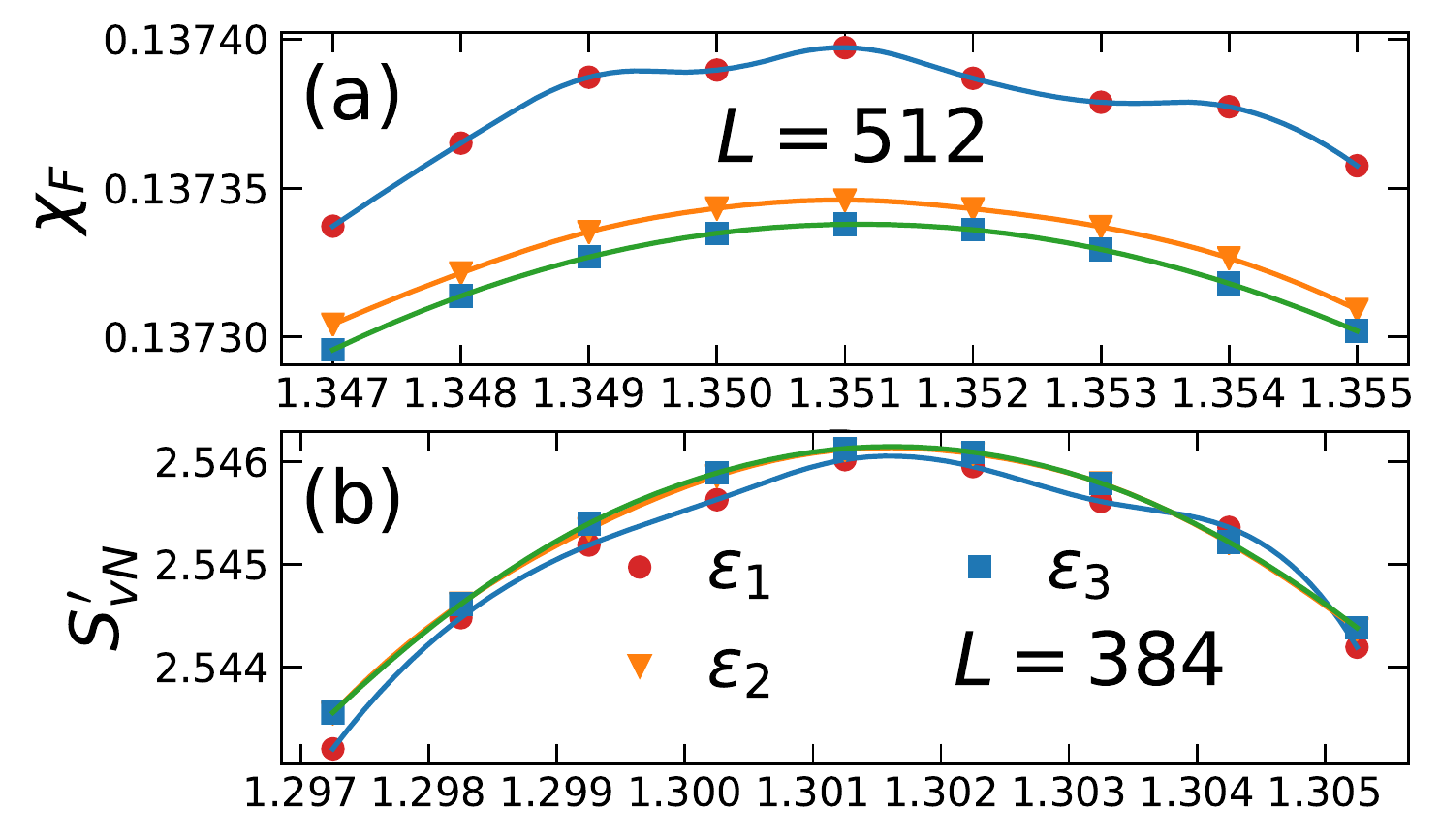}
    \caption{\label{fig:spfs3trunc}(a) The derivative of $S_{vN}$ for $L = 384$ and (b) the fidelity susceptibility for $L = 512$ as a function of $D$ around their peaks, respectively. The results are for Hamiltonian \eqref{eq:any-spin-ham-u} with $S = 2$ truncation. Three cases with different truncation errors $\epsilon_1 = 10^{-10}$, $\epsilon_2 = 10^{-11}$, and $\epsilon_3 = 10^{-12}$ are presented here.
    }
  \end{figure}

To find the peak heights and the peak positions for $\chi_F$ and $S^{\prime}_{vN}$, we apply a spline interpolation on data sets with $\Delta D = 10^{-3}$. Figure~\ref{fig:spfs3trunc}(a) shows that the variation of the values of $\chi_F$ inside a $0.01$ interval around the peak is of order $10^{-5}$ for $L = 512$, thus the results from DMRG with $\epsilon_1 = 10^{-10}$ are not accurate enough and have fluctuations. The results for $\epsilon_2 = 10^{-11}$ and $\epsilon_3 = 10^{-12}$ are smooth, and there is a small discrepancy of order $10^{-6}$ between them. Similar behaviors can be seen for $S^{\prime}_{vN}$ in Fig.~\ref{fig:spfs3trunc}(b), where the change in $S^{\prime}_{vN}$ inside a $0.01$ interval is of order $10^{-3}$. The error in $S^{\prime}_{vN}$ for $\epsilon_1$ is still big and results in fluctuations. The discrepancy between the results for $\epsilon_2$ and those for $\epsilon_3$ is of order $10^{-5}$ so invisible.

Because the logarithmic corrections result in slow convergence of observables in finite-size systems, it is necessary to have accurate data for finite $L$ to avoid large error propagation in the extrapolation procedure. We use a truncation error $\epsilon_3 = 10^{-12}$ in DMRG calculations for Hamiltonian \eqref{eq:any-spin-ham-u} with $S = 1$ and $S = 2$. Among the cases we calculated, the maximal bond dimension of MPS is $D_{\rm{bond}} = 962$ for $S = 1, L = 1024, D = 0.775$. 
In other calculations for Hamiltonian \eqref{eq:any-spin-ham-u} with $S = 3, 4$ and Hamiltonian \eqref{eq:any-spin-ham-ladder} with $S = 2, 3, 4, 5$, we use a truncation error $\epsilon_2 = 10^{-11}$, and the largest bond dimension of MPS is $D_{\rm{bond}} = 580$ for Hamiltonian \eqref{eq:any-spin-ham-ladder} with $S = 2, L = 512, D = 1.148$.

\section{Results}\label{sec:results}
We discuss the numerical results in this section. We have obtained accurate values of the phase-transition points from LS in previous work (see Table \ref{tableYcSsfs} or Ref. \cite{PhysRevB.103.245137}), which are used as references to describe the plots and check the accuracy of the results from $\chi_F$ and $S^{\prime}_{vN}$. 
\subsection{General remarks} \label{subsec:genremark}
\begin{figure}
  \centering
    \includegraphics[width=0.48\textwidth]{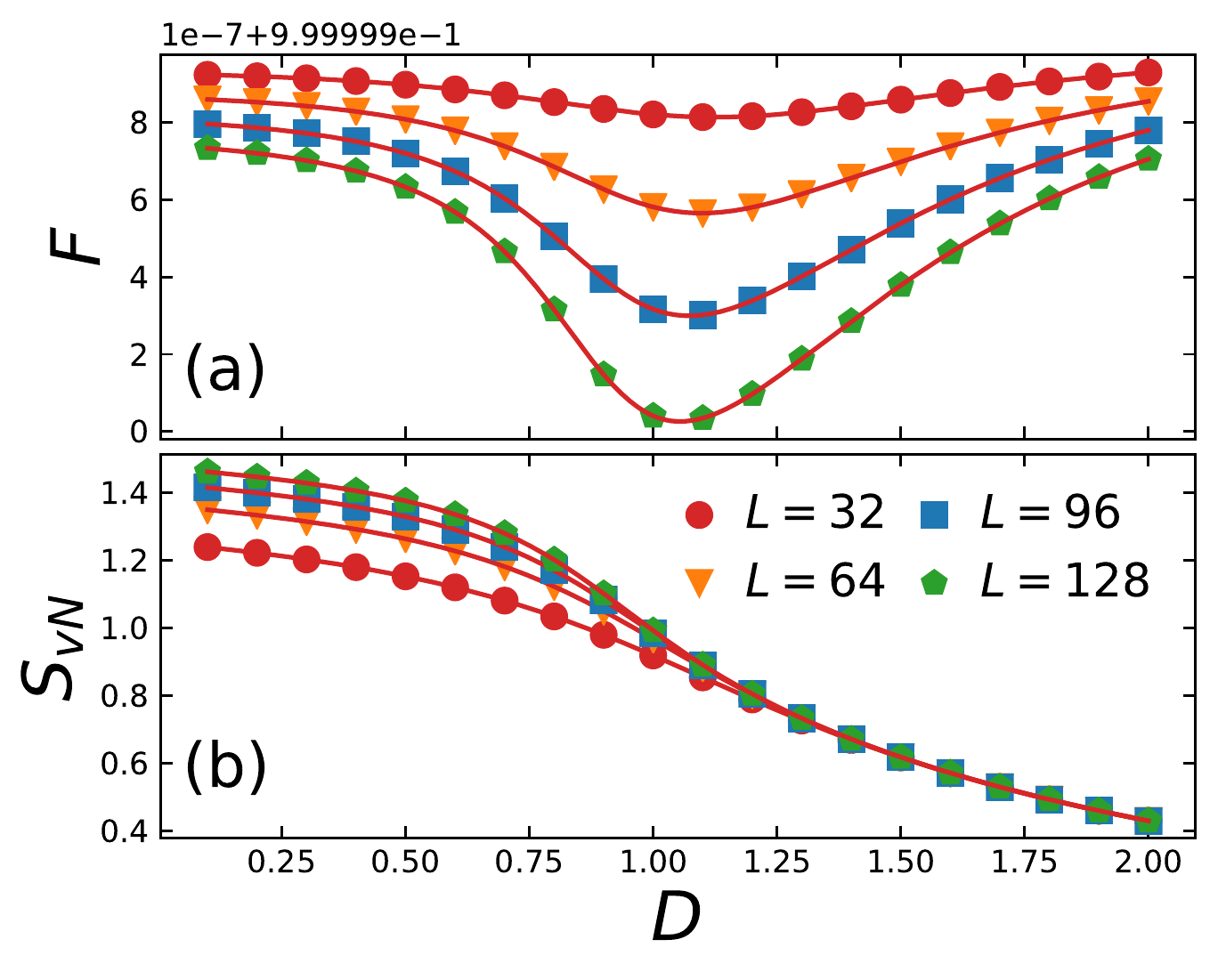}
    \caption{\label{fig:s1fidsvn}(a) Fidelity and (b) entanglement entropy as a function of $D$ for $S = 1$, $L = 32, 64, 96, 128$.
    }
  \end{figure}

\begin{figure}
  \centering
    \includegraphics[width=0.48\textwidth]{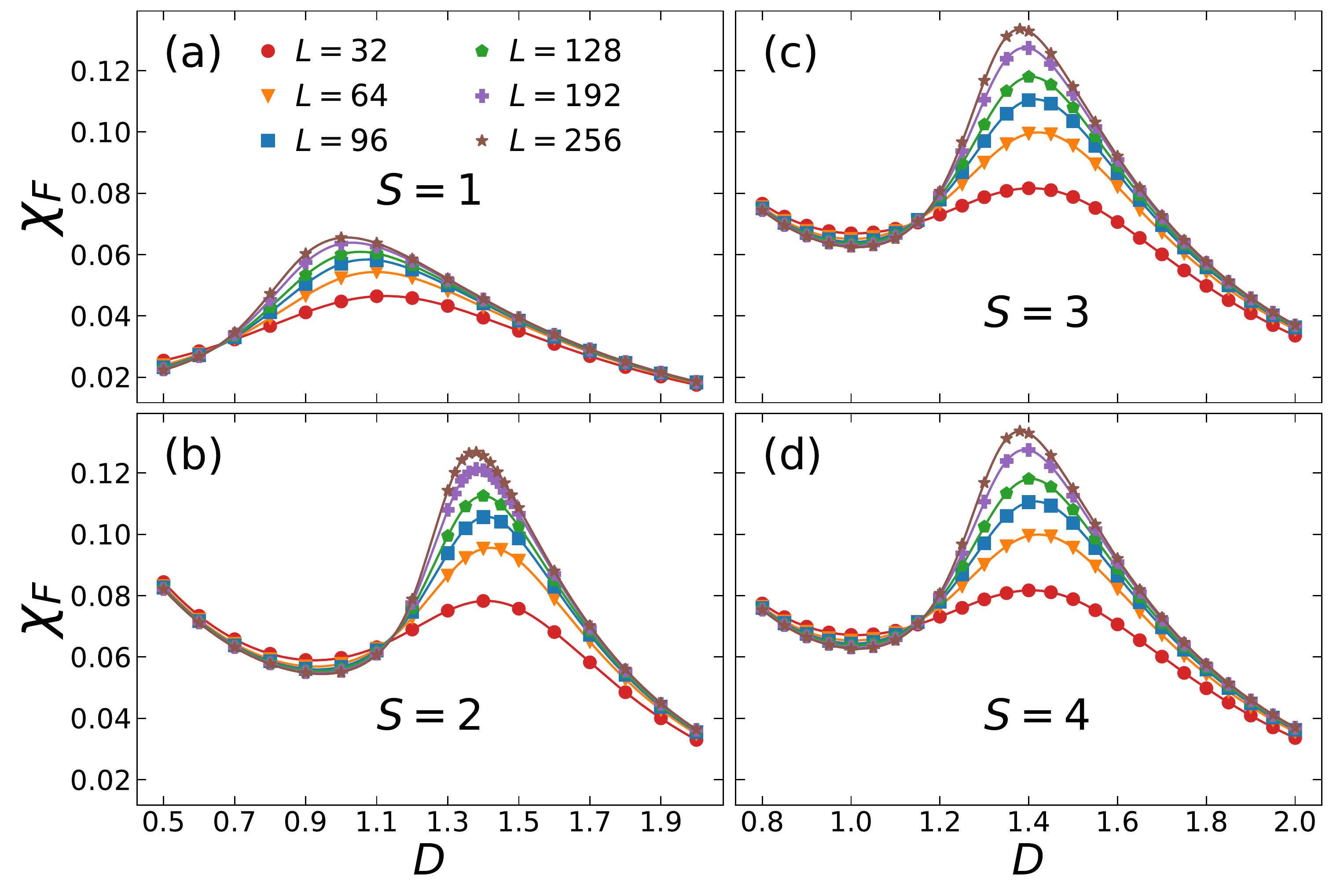}
    \caption{\label{fig:s1234fidsus}$\chi_F$ as a function of $D$ for (a) $S = 1$, (b) $S = 2$, (c) $S = 3$, and (d) $S = 4$. The results are for  Hamiltonian \eqref{eq:any-spin-ham-u} with $L = 32, 64, 96, 128, 192, 256$.
    }
  \end{figure}

We first present the behavior of $F$, $\chi_F$, $S_{vN}$, and $S^{\prime}_{vN}$ as functions of the coupling constant $D$. As we discussed before, the fidelity will not have a drastic drop for our Hamiltonians that have IOQPTs without symmetry breaking, and the entanglement entropy will keep increasing as we decrease the value of $D$. Figure~\ref{fig:s1fidsvn} confirms our expectations. The results are for Hamiltonian \eqref{eq:any-spin-ham-u} with $S = 1$. In Fig.~\ref{fig:s1fidsvn}(a), we see that the fidelity does have a drop. The magnitude of the drop is only of order $10^{-7}$ for $L \lesssim 100$, and increases almost linearly with $L$. Based on the definition of $\chi_F$ in Eq.~\eqref{eq:fidsus}, the fidelity susceptibility will not diverge in the thermodynamic limit, and is of order $10^{-2}$, which is confirmed in Fig.~\ref{fig:s1234fidsus}. Figure~\ref{fig:s1fidsvn}(b) shows that the entanglement entropy is independent with the size of the system for large $D$, indicating that the large-$D$ phase is gapped. For small $D$, $S_{vN}$ increases as we increase $L$, and one can see that the increment in $S_{vN}$ by doubling the size of the system is almost the same ($\sim 0.11$), consistent with the logarithmic scaling in Eq.~\eqref{eq:eescaling}. Note that there is a local extreme in the entanglement entropy at the BKT transition of the spin-$1/2$ $XXZ$ chain \cite{PhysRevA.81.032334}, while there is no such phenomenon to signal the IOQPTs in our models. Similar behaviors are seen in Hamiltonians \eqref{eq:any-spin-ham-u} and \eqref{eq:any-spin-ham-ladder} with any $S$ truncations (not shown here).

\begin{figure}
  \centering
    \includegraphics[width=0.48\textwidth]{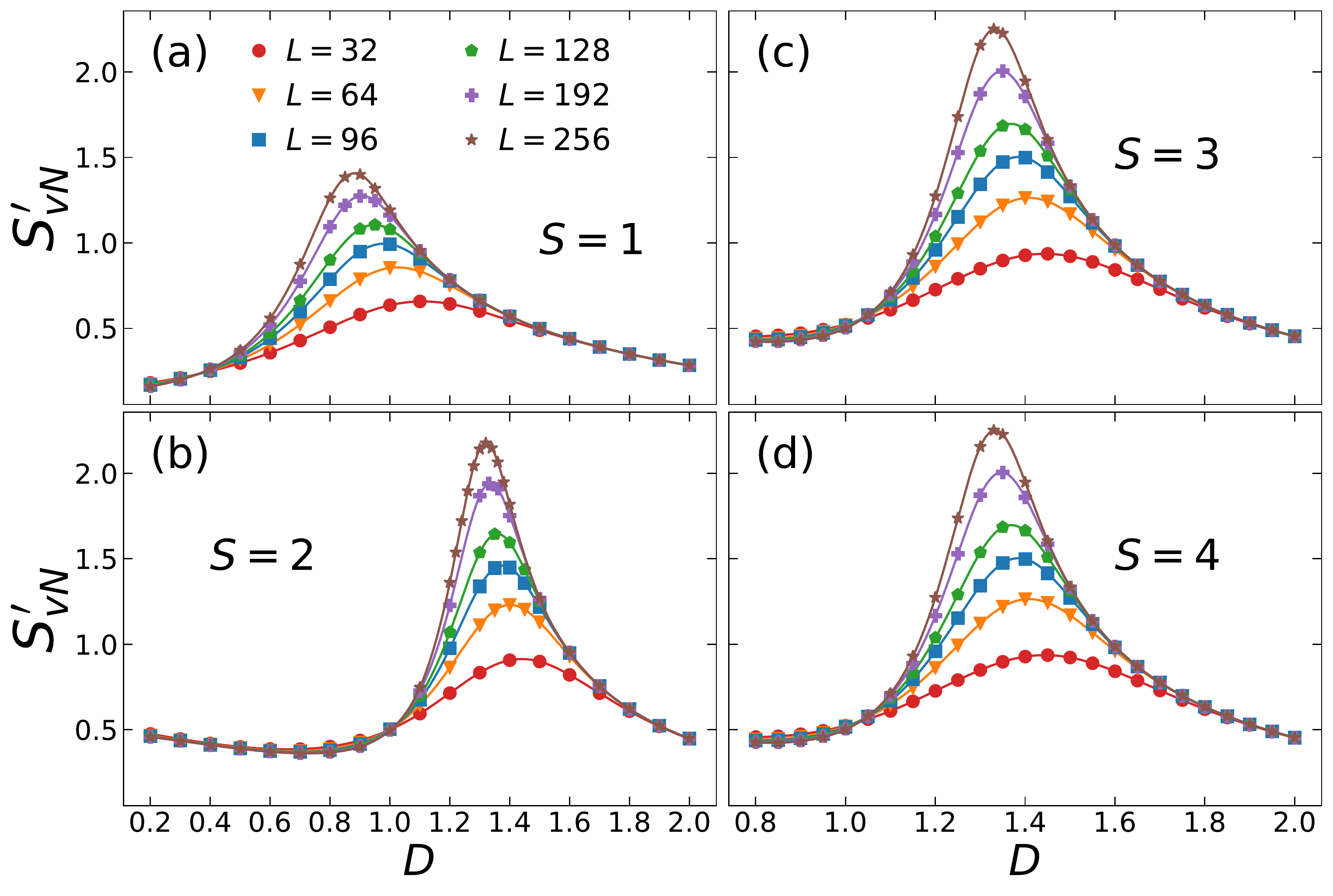}
    \caption{\label{fig:s1234dsdy}Same as Fig.~\ref{fig:s1234fidsus}, but for $S^{\prime}_{vN}$.
    }
  \end{figure}
  
Since the change in the fidelity is very small, it is difficult to measure in any quantum simulation experiment. The expression of $\chi_F$ in Eq.~\eqref{eq:fsde0}, however, is related to the spectral function that can be measured experimentally \cite{SJGu2014,You_2015}. The entanglement entropy does not have a local extreme to signal a QPT, but the derivative of $S_{vN}$ obviously does. We present $\chi_F$ and $S^{\prime}_{vN}$ for Hamiltonian \eqref{eq:any-spin-ham-u} with $S = 1, 2, 3, 4$ as functions of $D$ in Fig.~\ref{fig:s1234fidsus} and Fig.~\ref{fig:s1234dsdy}, respectively. We see that the results for $S = 1$ is very different from those for $S \ge 2$, and the difference between the results for $S = 3$ and those for $S = 4$ is invisible. One can expect that both $\chi_F$ and $S^{\prime}_{vN}$ for finite size systems converge exponentially with $S$. For all the cases, the peaks of $\chi_F$ and $S^{\prime}_{vN}$ move slowly with increasing $L$. According to Eq.~\eqref{eq:corrlens12}, the peaks should move to their thermodynamic positions with a leading behavior of $1/\ln(L)$ for $S = 1$ and $1/\ln^2(L)$ for $S \ge 2$. Because $1/\ln^2(L)$ decreases more slowly with increasing $L$ than $1/\ln(L)$ does, so the peaks for $S \ge 2$ move slower than that for $S = 1$ does. But $1/\ln^2(L)$ term is much smaller than $1/\ln(L)$ term, so for the same finite $L$, the peak position of $S = 1$ is farther away from the phase-transition point than those of $S \ge 2$. We check these scaling behaviors in Fig.~\ref{fig:s12yccompare}. Then we see that the peak height of $\chi_F$ ($\chi^*_{F}$) increases very slow and is not likely to diverge for all cases. The value of $\chi^*_{F}$ for $S = 1$ grows more slowly than those for $S \ge 2$, indicating a different convergent behavior. The peak height of $S^{\prime}_{vN}$ ($S^{\prime *}_{vN}$) tends to diverge for all the cases, but the one for $S = 1$ again diverges slower than others.

Notice that in all plots shown here, there exists a crossing point for curves between different system sizes. The crossing points are all close to the phase-transition points, except for $\chi_F$ at $S = 1$, which may be due to large finite-size effects (discussed further in Sec. \ref{subsec:crosspoint}). Assuming both $\chi_F$ and $S^{\prime}_{vN}$ are single-valued functions of $D$, the peaks will never go to the left side of the crossing points. If the crossing point is larger than the phase-transition point, the peak position is also larger than the phase-transition point. This criteria can be used to check the FSSs of the peaks of $\chi_F$ and $S^{\prime}_{vN}$. In the following, we first study the FSSs of the peaks of $\chi_F$ and $S^{\prime}_{vN}$, and then discuss the FSS of the crossing point to crosscheck the results. The main conclusion is that the peak position of $\chi_F$ is inside the gapped phase and larger than the IOQPT point, while the peak position of $S^{\prime}_{vN}$ is at the IOQPT point for all $S$ truncations. But both the FSS of the peak height of $\chi_F$ and that of $S^{\prime}_{vN}$ can be used to differentiate between the IOG transition and BKT transitions.

\subsection{Finite-size scaling of peaks} \label{subsec:fss}
\begin{figure}
  \centering
    \includegraphics[width=0.48\textwidth]{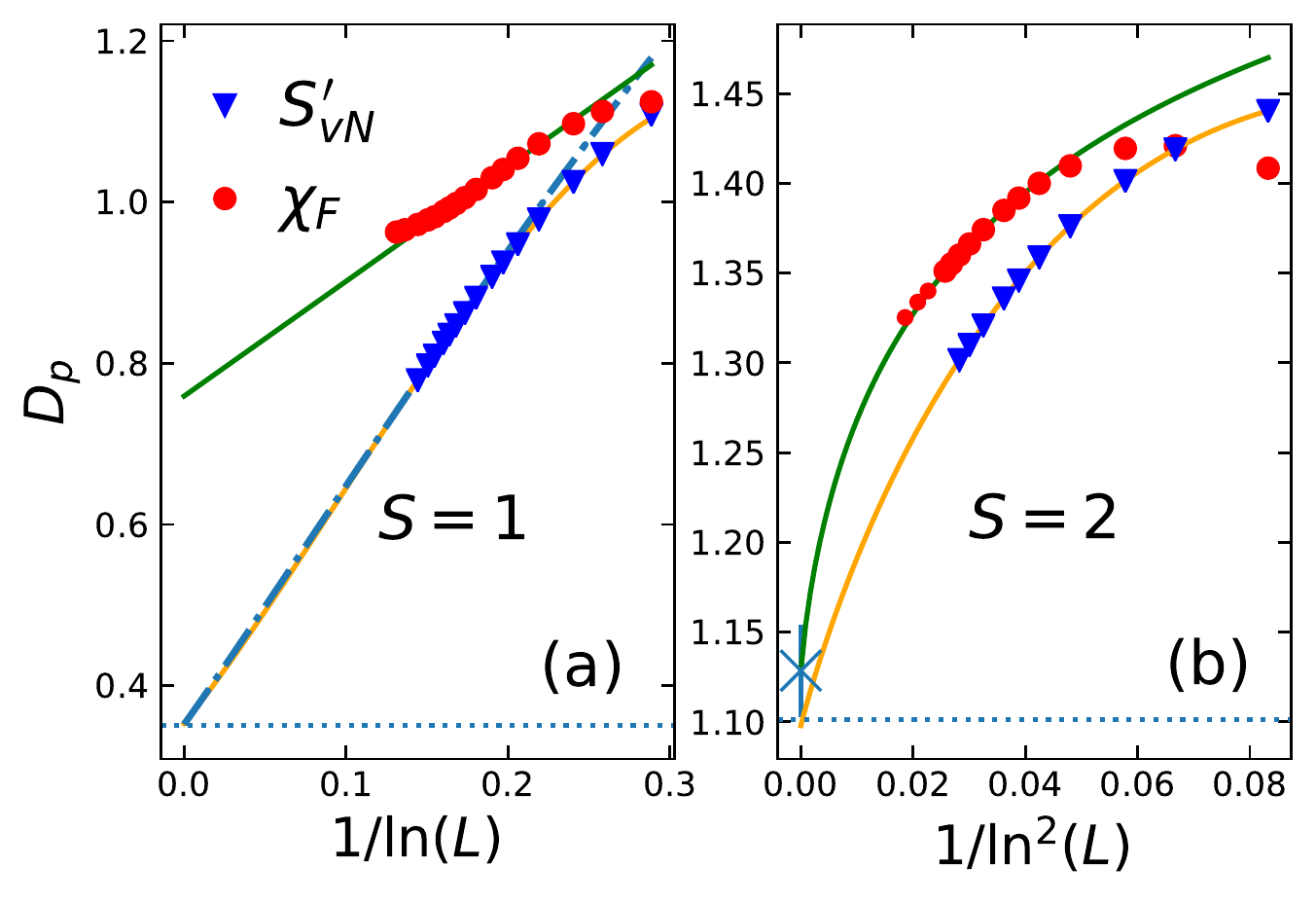}
    \caption{\label{fig:s12yccompare} Extrapolation procedures for the phase transition point $D_c$ at (a) $S = 1$ truncation and (b) $S = 2$ truncation. The extrapolations are performed with peak positions of $\chi_F$ (circles), and peak positions of $S^{\prime}_{vN}$ (triangles). For $S = 1$, solid lines on the symbols are the curve fitting $C+A/\ln(L)$ for $\chi_F$, and curve fitting with Eq.~\eqref{eq:delDvslnLs1} for $S^{\prime}_{vN}$, respectively. The dash-dot line on triangles fits two data points for the largest $L = 768, 1024$ with Eq.~\eqref{eq:delDvslnLs1} setting $d=e=0$. For $S = 2$, solid lines are fitting with Eq.~\eqref{eq:fspeakfss} for $\chi_F$, and fitting with Eq.~\eqref{eq:delDvslnLs2} for $S^{\prime}_{vN}$. The extrapolated $D_c$ for $S = 1$ is $0.759(3)$ for $\chi_F$, $0.353(7)$ (solid line) and $0.3523$ (dash-dot line) for $S^{\prime}_{vN}$, respectively. The extrapolated $D_c$ for $S = 2$ is $1.129(26)$ and $1.0979(3)$, respectively. The dashed lines are results from level spectroscopy in Ref. \cite{PhysRevB.103.245137}. The results are for Hamiltonian \eqref{eq:any-spin-ham-u}. The smaller red circles in (b) are results for $L = 768, 1024, 1536$, which are not used in the curve fitting and are used to show that the best fitting function is below the true values of $D_p$ for larger $L$.
    }
  \end{figure}

As shown in Fig.~\ref{fig:spfs3trunc}, we perform a spline interpolation inside a $0.01$ interval for an equidistant data set with $\Delta D = 10^{-3}$ to find the peak positions $D_p$ and the peak heights $\chi^*_F$ and $S^{\prime *}_{vN}$. We first discuss the results for Hamiltonian \eqref{eq:any-spin-ham-u} with $S = 1, 2$, where we have the most accurate data from DMRG with truncation error $\epsilon_3 = 10^{-12}$. Figure~\ref{fig:s12yccompare} depicts the results for the peak positions and the procedures of extrapolations to $L \rightarrow \infty$. The minimal system size we calculated is $L = 32$ for all cases, and the maximal system sizes are $L = 2048$ for $\chi_F$ at $S = 1$, $L = 1024$ for $S^{\prime}_{vN}$ at $S = 1$, $L = 512$ for $\chi_F$ at $S = 2$, and $L = 384$ for $S^{\prime}_{vN}$ at $S = 2$. In Fig.~\ref{fig:s12yccompare}(a), we see that $D_p$ of $\chi_F$ at $S = 1$ is linear with $1/\ln(L)$ but only for intermediate system sizes, where we fit the data with a linear function of $1 / \ln(L)$ for $160 \le L \le 512$, and find the extrapolated value $0.759(3)$, which is far from the IOG transition point $D_c = 0.35067$ from LS \cite{PhysRevB.103.245137}. Moreover, the value of $D_p$ starts to decrease slower than $1/\ln(L)$ for system sizes larger than $L = 512$, so the peak position of $\chi_F$ must be larger than $0.759$ and does not signal the IOG point for $S = 1$. In Fig.~\ref{fig:s12yccompare}(b) for $S = 2$, we use Eq.~\eqref{eq:fspeakfss} proposed in Ref. \cite{PhysRevB.91.014418} to fit the peak positions for $160 \le L \le 512$, and find the extrapolated $D_c = 1.129(26)$, which is close to the result $D_c = 1.1013$ from LS \cite{PhysRevB.103.245137}. So we have confirmed our speculation in Sec.~\ref{subsec:modelgsfs} that we can extrapolate approximate values of the BKT points using data for intermediate system sizes. However, the peak of $\chi_F$ is not singular, which may be shifted by other source of contributions, thus cannot single out the BKT point accurately. Using a smaller truncation error $\epsilon_1 = 10^{-10}$, we can determine the value of $D_p$ with an error less than $10^{-3}$ for larger system sizes. We find that $D_p = 1.340, 1.334, 1.325$ for $L = 768, 1024, 1536$ from DMRG, respectively. The three data points are shown in Fig.~\ref{fig:s12yccompare}(b) as small red circles. The extrapolated values from the curve fit are $1.3390, 1.3310, 1.3205$. Although not as pronounced as $S = 1$ case, the extrapolated value is also smaller than the true value and the error increases with $L$. So the FSS [$A/\ln^2(BL)$] deduced from the divergent behavior of the correlation length is not true for large systems. These observations do not contradict the results in other works. For example, Ref. \cite{PhysRevB.100.094428} applies this scaling to clock models and successfully find the BKT points, where only system sizes $L \le 144$ are used. Now we use data sets for $L \le 512$ to extrapolate the values of BKT points for all the other cases. In table \ref{tableYcSsfs}, we list the extrapolated values of $D_c$ from $\chi_F$ for Hamiltonian \eqref{eq:any-spin-ham-u} with $S = 1,2,3,4$ and Hamiltonian \eqref{eq:any-spin-ham-ladder} with $S = 1,2,3,4,5$, and compare them with those from LS. We see that all the extrapolated values of BKT points deviate from the true $D_c$ by an amount of order $10^{-2}$. 
\begin{table}
\caption{\label{tableYcSsfs}Extrapolated values of the phase-transition points $D_c$ from $\chi_F$ for Hamiltonians \eqref{eq:any-spin-ham-u} and \eqref{eq:any-spin-ham-ladder} with different $S$. Data points for intermediate system sizes $160 \le L \le 512$ are used in the extrapolations. The results from level spectroscopy (LS) \cite{PhysRevB.103.245137} are also shown for comparisons.}
    \begin{tabular} {p{1.4cm} p{1.6cm} p{1.6cm}  p{1.6cm}  p{1.6cm}}
    \hline
    \hline
        &$\hat{H}_U, \chi_F$ &$\hat{H}_U$, LS &$\hat{H}_S, \chi_F$ &$\hat{H}_S$, LS  \\  \hline
    $S = 1$ &0.756(6) &0.3507 &0.756(6) &0.3507 \\  \hline
    $S = 2$ &1.129(26) &1.1013 &1.02(4) &0.9322   \\  \hline
    $S = 3$ &1.106(17) &1.1256 &0.95(5) &1.0331  \\  \hline
    $S = 4$ &1.08(18) &1.1259 &1.089(8) &1.0710  \\  \hline
    $S = 5$ & & &1.11(3) &1.0895
    \\  \hline \hline
    \end{tabular}  
\end{table}

The results for $S^{\prime}_{vN}$ are much more accurate. In Fig.~\ref{fig:s12yccompare}(a), we see that $D_p$ for $S = 1$ becomes linear with $1/\ln(L)$ quickly. We can use the leading scaling to fit the values of $D_p$ for the biggest two system sizes $L = 768, 1024$ and obtain the extrapolated $D_c = 0.357$, which has a difference only of $0.006$ from the result from LS. We can improve the result by adding subleading corrections. We first consider the a correction term from $(D-D_c)^{-1/2}$ in Eq.~\eqref{eq:corrlens12}, and use the first line of Eq.~\eqref{eq:delDvslnLs1} to fit the two data points for $L = 768, 1024$. We obtain the extrapolated $D_c = 0.3523$, much closer to the result from LS. Higher-order corrections are complicated and make fitting procedure unstable. We consider two more correction terms proportional to $1/\ln^2(L)$ and $1/\ln^3(L)$, and use Eq.~\eqref{eq:delDvslnLs1} to fit data points for $L = 96, 128, ..., 1024$ and find $D_c = 0.353(7)$ and $b_1 = 2.49(11)$, which are consistent with the results [$D_c = 0.3512(10), b_1 = 2.501(13)$] from gap scaling \cite{PhysRevB.103.245137}. The extrapolation procedure for $S = 2$ is shown in Fig.~\ref{fig:s12yccompare}(b). The leading term $1/\ln^2(L)$ has not dominate the scaling for the maximal $L = 384$ we calculated, but the higher-order corrections with higher powers of $1/\ln(L)$ can help improve the extrapolated results. We use Eq.~\eqref{eq:delDvslnLs2} to fit the data points for $L = 32, 48, ..., 384$ and find that $D_c = 1.0979(3)$ and $b_2 = 3.597(5)$. In Table \ref{tableYcSsdsdy}, we summarize the extrapolated values of $D_c$ from $S^{\prime}_{vN}$, where one can see that all the results are close to those from LS with differences only of order $10^{-3}$. 
\begin{table}
\caption{\label{tableYcSsdsdy}Extrapolated values of the phase-transition points $D_c$ from $S^{\prime}_{vN}$ for different $S$.}
    \begin{tabular} {p{2.5cm} p{2.5cm}  p{2.5cm}}
    \hline
    \hline
        &$\hat{H}_U, S^{\prime}_{vN}$ &$\hat{H}_S, S^{\prime}_{vN}$  \\  \hline
    $S = 1$ &0.353(7) &0.353(7)  \\  \hline
    $S = 2$ &1.0979(3) &0.9401(14)   \\  \hline
    $S = 3$ &1.120(2) &1.038(3)  \\  \hline
    $S = 4$ &1.122(3) &1.069(6) \\  \hline
    $S = 5$ & &1.085(9)
    \\  \hline \hline
    \end{tabular}  
\end{table}
The values of $b_S$ are summarized in Table \ref{tablebSSsdsdy}, which are close to those from gap scaling (GS) \cite{PhysRevB.103.245137}. Because the method of gap scaling does not include higher-order corrections, we believe the results here are more accurate, which can be crosschecked with the value of central charge $c = 1$ (see below).
\begin{table}
\caption{\label{tablebSSsdsdy}Values of $b_S$ from $S^{\prime}_{vN}$ for different $S$. The results from gap scaling (GS) \cite{PhysRevB.103.245137} are also shown for comparisons.}
    \begin{tabular} {p{1.4cm} p{1.6cm} p{1.6cm}  p{1.6cm}  p{1.6cm}}
    \hline
    \hline
        &$\hat{H}_U, S^{\prime}_{vN}$ &$\hat{H}_U$, GS &$\hat{H}_S, S^{\prime}_{vN}$ &$\hat{H}_S$, GS  \\  \hline
    $S = 1$ &2.49(11) &2.501(13) &2.49(11) &2.501(13)  \\  \hline
    $S = 2$ &3.597(5) &3.2553(21) &3.767(24) &3.647(4)  \\  \hline
    $S = 3$ &3.52(4) &3.110(5) &3.53(5) &3.367(2)  \\  \hline
    $S = 4$ &3.49(5) &3.117(5) &3.57(10) &3.281(3)  \\  \hline
    $S = 5$ & & &3.58(14) &3.25(1) 
    \\  \hline \hline
    \end{tabular}  
\end{table}

Before going to the discussion of the peak heights, we add a side remark for the shift of the peak position of $\chi_F$ away from the phase-transition points. Reference~\cite{PhysRevB.100.081108} asserts that the shift of the peak position of $\chi_F$ is $b_S^2/36$ for BKT transitions, which is $0.36$ for $S = 2$. But in Fig.~\ref{fig:s12yccompare}(b), our $D_p$ is smaller than $D_c+0.36 = 1.46$ for all system sizes. This is because our $b_S$ is large, and the result in Ref.~\cite{PhysRevB.100.081108} is more accurate for smaller $b_S$. Following the derivations in Ref.~\cite{PhysRevB.100.081108}, one can obtain the shifted peak position of $\chi_F$ for the IOG transition is $D_c + 2b_1/9 \approx 0.91$, close to the value of $D_p = 0.96$ for $L = 2048$. Algorithms for infinite-size systems are needed to check this, which is beyond the scope of this work.

\begin{figure}
  \centering
    \includegraphics[width=0.48\textwidth]{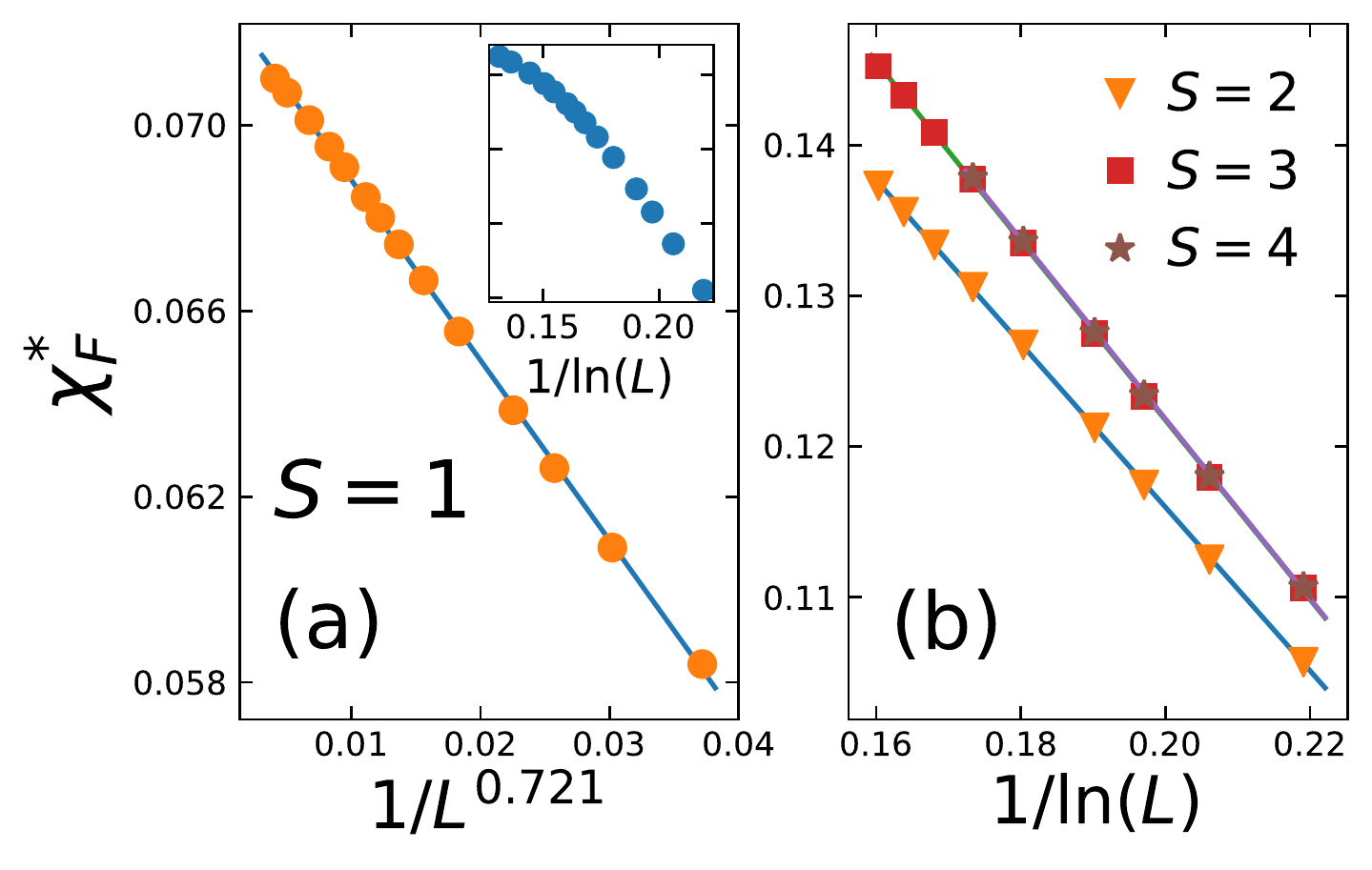}
    \caption{\label{fig:s1234fspeakscaling}The peak height of $\chi_F$ as a function of (a) $1/L^{0.721}$ for $S = 1$ and (b) $1/\ln(L)$ for $S = 2, 3, 4$, respectively. The solid lines are linear fits. The markers for $S = 3$ and $S = 4$ are on top of each other. The inset of (a) shows $\chi^*_F$ for $S = 1$ as a function of $1/\ln(L)$. The results are for Hamiltonian \eqref{eq:any-spin-ham-u}.
    }
  \end{figure}
  
\begin{figure}
  \centering
    \includegraphics[width=0.48\textwidth]{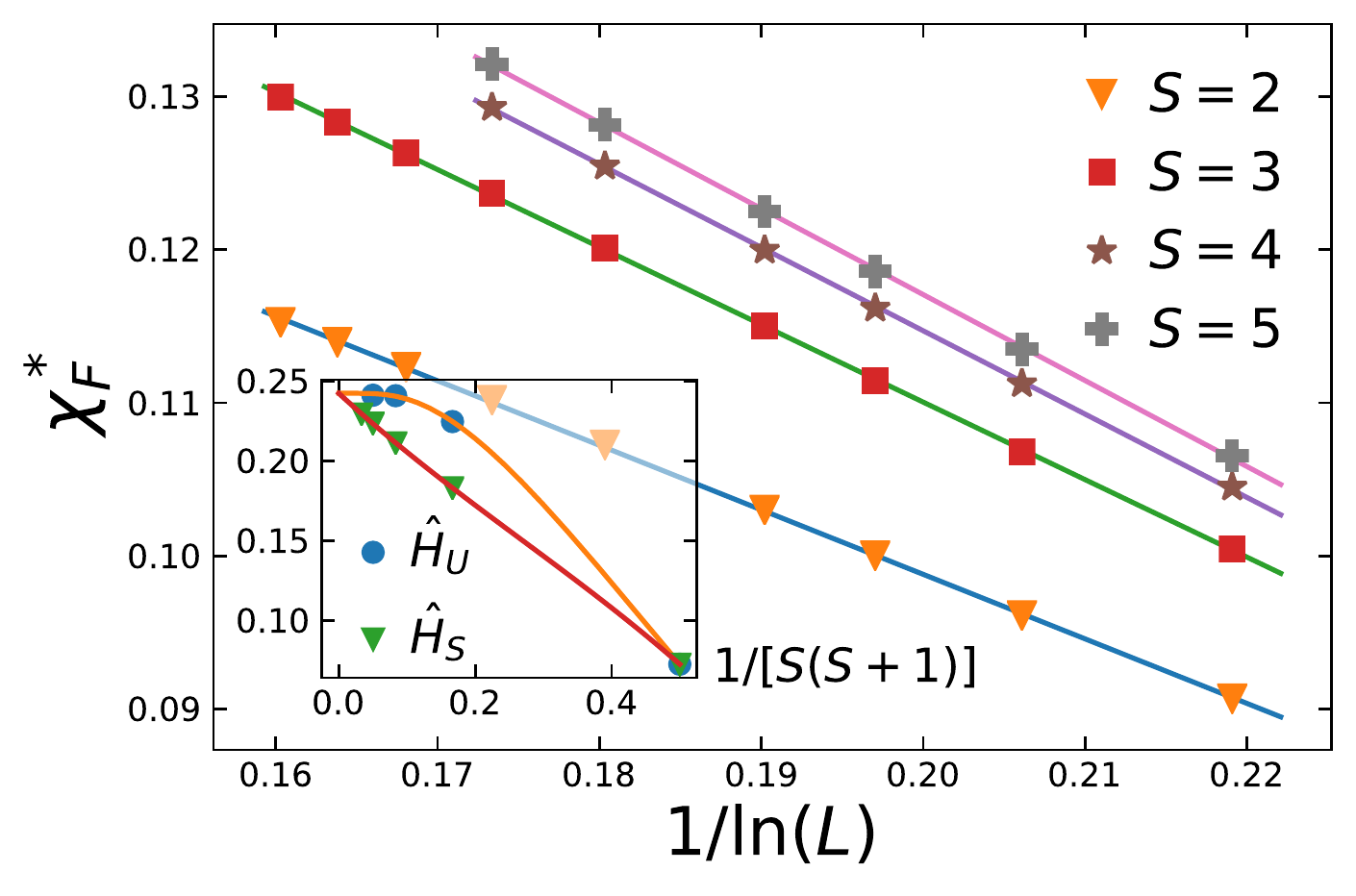}
    \caption{\label{fig:sop2345fspeakscaling}Same as Fig.~\ref{fig:s1234fspeakscaling}, but for Hamiltonian \eqref{eq:any-spin-ham-ladder}. The inset presents the extrapolated thermodynamic values of $\chi^*_F$ as a function of $1/[S(S+1)]$. The results for Hamiltonian \eqref{eq:any-spin-ham-u} from Fig.~\ref{fig:s1234fspeakscaling} are fit with an exponential convergence function of $S$, while those for Hamiltonian \eqref{eq:any-spin-ham-ladder} are fit with a polynomial function of $1/[S(S+1)]$.
    }
  \end{figure}
  
We next discuss the scaling of the peak height of $\chi_F$ and $S^{\prime}_{vN}$. In Fig.~\ref{fig:s1234fspeakscaling}, we present the values of $\chi^{*}_F$ for Hamiltonian \eqref{eq:any-spin-ham-u} as a function of $L$. For BKT transitions ($S \ge 2$), $\chi^*_F$ is expected to scale linearly with $1/\ln(L)$ \cite{PhysRevB.91.014418}, which is confirmed in Fig.~\ref{fig:s1234fspeakscaling}(b). However, for $S = 1$, the inset of Fig.~\ref{fig:s1234fspeakscaling}(a) shows that $\chi^*_F$ is not linear with $1/\ln(L)$. We do not think $\chi^*_F$ scales polynomially with $1/L$, either, because the coefficients of $1/L^p$ ($p \ge 2$) are unreasonably large in the curve fit. We fit the data with a power-law function of $1/L$ and find that $\chi^*_F \sim 1/L^{0.721(15)}$ for $S = 1$. The results for Hamiltonian \eqref{eq:any-spin-ham-ladder} with $S = 2, 3, 4, 5$ are shown in Fig.~\ref{fig:sop2345fspeakscaling}, where the values of $\chi^*_F$ are all linear with $1/\ln(L)$. So the scaling of the peak height of $\chi_F$ can differentiate between IOG transitions and BKT transitions. Notice that the values of $\chi^*_F$ for Hamiltonian \eqref{eq:any-spin-ham-u} converge quickly with $S$ and have invisible difference between $S = 3$ and $S = 4$, while those for Hamiltonian \eqref{eq:any-spin-ham-ladder} converge much slower and have clear difference between $S = 4$ and $S = 5$. These phenomena are consistent with the exponential convergence of energy gap and phase-transition points discussed in Ref. \cite{PhysRevB.103.245137}. We fit the extrapolated values of $\chi^*_F$ for Hamiltonian \eqref{eq:any-spin-ham-u} with an exponential convergence function of $S$, and fit those for Hamiltonian \eqref{eq:any-spin-ham-ladder} with a polynomial function of $1/[S(S+1)]$, and find that the values of the peak height of $\chi_F$ for $L \rightarrow \infty, S \rightarrow \infty$ are $0.2423(7)$ and $0.2422(7)$, respectively, which are the same within uncertainties as expected.  

\begin{figure}
  \centering
    \includegraphics[width=0.48\textwidth]{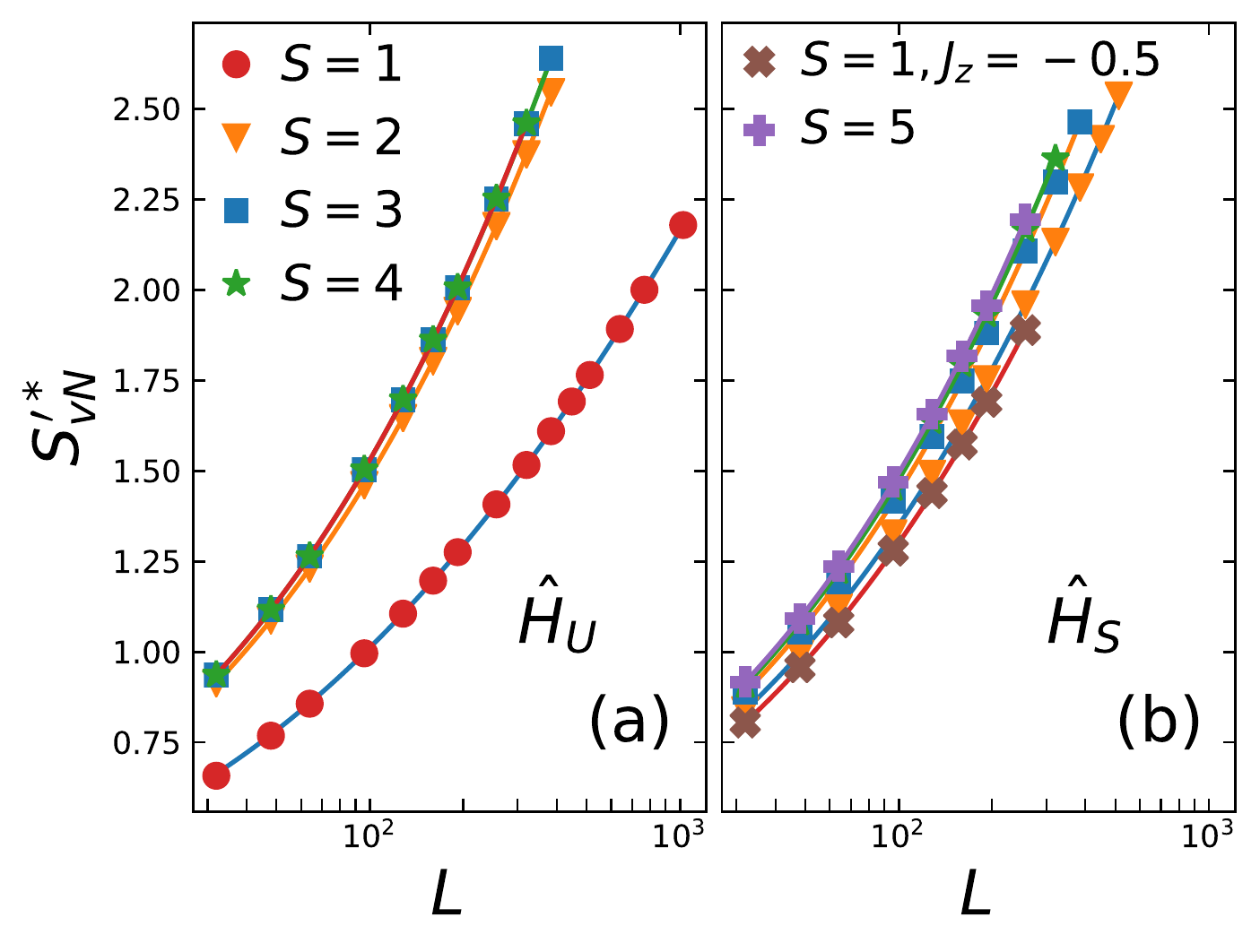}
    \caption{\label{fig:s12345dsdypeakscaling}The peak height of $S^{\prime}_{vN}$ as a function of $L$ for (a) Hamiltonian \eqref{eq:any-spin-ham-u} with $S = 1, 2, 3, 4$ and (b) Hamiltonian \eqref{eq:any-spin-ham-ladder} with $S = 2, 3, 4, 5$. The solid lines are fits with Eq.~\eqref{eq:dsdys1} for $S = 1$ and with Eq.~\eqref{eq:dsdys2} for others. The result for a true BKT transition in Hamiltonian \ref{eq:any-spin-ham-u} with $S = 1$ plus $J_z = -0.5$ term is also displayed in (b). Markers for $S = 3$ and $S = 4$ are on top of each other in (a), and those for $S = 4$ and $S = 5$ are on top of each other in (b).
    }
  \end{figure}
  
Figure~\ref{fig:s12345dsdypeakscaling} depicts the the peak height of $S^{\prime}_{vN}$ as a function of $L$. The results for Hamiltonian \eqref{eq:any-spin-ham-u} are shown in Fig.~\ref{fig:s12345dsdypeakscaling}(a), and those for Hamiltonian \eqref{eq:any-spin-ham-ladder} are shown in Fig.~\ref{fig:s12345dsdypeakscaling}(b). We also present the result for Hamiltonian \eqref{eq:any-spin-ham-u} with $S = 1$ plus a nearest-neighbor-interaction term $\sum_l S^z_l S^z_{l+1}$ with a coupling constant $J_z = -0.5$, which also has a BKT transition \cite{PhysRevB.67.104401}, in Fig.~\ref{fig:s12345dsdypeakscaling}(b). Firstly, one can also see that $S^{\prime *}_{vN}$ for Hamiltonian \eqref{eq:any-spin-ham-u} converges faster with $S$ than that for Hamiltonian \eqref{eq:any-spin-ham-ladder} does. More importantly, the scaling of $S^{\prime *}_{vN}$ for $S = 1$ is slower than so obviously different from others. The plots for all the cases that have BKT transitions, including the one with $S = 1, J_z = -0.5$, are almost parallel with each other, indicating that they have the same leading scaling. We use Eq.~\eqref{eq:dsdys1} to fit the data for $S = 1$ and system sizes $L = 256, 320, \ldots, 1024$. For BKT cases, we take the scaling form in Eq.~\eqref{eq:dsdys2} to fit the data for system sizes starting from $L = 32$. The values of the power $b_S$ can be extracted and are summarized in Table \ref{tablepSSsdsdy}.
\begin{table}
\caption{\label{tablepSSsdsdy}Values of $p_S$ from $S^{\prime}_{vN}$ for different $S$.}
    \begin{tabular} {p{2.5cm}  p{2.5cm}  p{2.5cm}}
    \hline
    \hline
        &$\hat{H}_U$ &$\hat{H}_S$  \\  \hline
    $S = 1$ &2.000(16) &2.000(16) \\  \hline
    \makecell[l]{$S = 1$, \\ $J_z = -0.5$} &2.930(7) &2.930(7) \\  \hline
    $S = 2$ &3.037(4) &2.930(4) \\  \hline
    $S = 3$ &3.075(4) &3.011(6) \\  \hline
    $S = 4$ &3.076(7) &3.047(6) \\  \hline
    $S = 5$ & &3.079(10)
    \\  \hline \hline
    \end{tabular}  
\end{table}
For the IOG transition in Hamiltonian \eqref{eq:any-spin-ham-u} with $S = 1$, we obtain $p_1 = 2.000(16)$, so $S^{\prime *}_{vN}$ diverges as $\ln^2(L)$. For BKT cases, the best-fit values of $p_S$ are all close to three with differences only of order $10^{-2}$, thus we have confirmed that $S^{\prime *}_{vN}$ diverges as $\ln^3(L)$ for BKT transitions.

To further support the validity of our results, we crosscheck the value of the central charge $c = 1$. As discussed in Sec. \ref{subsec:modelgsee}, $a_S$, $b_S$, and the central charge $c$ can be related by Eq.~\eqref{eq:a1bcrelation} for $S = 1$ and Eq.~\eqref{eq:a2bcrelation} for $S \ge 2$. We list the values of $a_S$ in Table \ref{tableaSSsdsdy}.
\begin{table}
\caption{\label{tableaSSsdsdy}Values of $a_S$ from $S^{\prime}_{vN}$ for different $S$.}
    \begin{tabular} {p{2.5cm}  p{2.5cm}  p{2.5cm}}
    \hline
    \hline
        &$\hat{H}_U$ &$\hat{H}_S$  \\  \hline
    $S = 1$ &0.064(4) &0.064(3) \\  \hline
    $S = 2$ &0.00648(8) &0.0062(9) \\  \hline
    $S = 3$ &0.00622(8) &0.00668(12) \\  \hline
    $S = 4$ &0.00621(13) &0.00631(11) \\  \hline
    $S = 5$ & &0.00590(17)
    \\  \hline \hline
    \end{tabular}  
\end{table}
Using the results in Table \ref{tablebSSsdsdy} and \ref{tableaSSsdsdy}, we calculate the values of central charge $c$ and put them in Table \ref{tableCentralCSsdsdy}.
\begin{table}
\caption{\label{tableCentralCSsdsdy}Values of central charge $c$ from $S^{\prime}_{vN}$ for different $S$.}
    \begin{tabular} {p{2.5cm}  p{2.5cm}  p{2.5cm}}
    \hline
    \hline
        &$\hat{H}_U$ &$\hat{H}_S$  \\  \hline
    $S = 1$ &0.96(10) &0.96(10) \\  \hline
    $S = 2$ &1.006(15) &1.06(17) \\  \hline
    $S = 3$ &0.92(3) &1.00(5) \\  \hline
    $S = 4$ &0.91(5) &0.97(7) \\  \hline
    $S = 5$ & &0.91(10)
    \\  \hline \hline
    \end{tabular}  
\end{table}
One can see that all the calculated values of the central charge are close to the expected value $c = 1$. In particular, for Hamiltonian \eqref{eq:any-spin-ham-u} with $S = 2$, where a truncation error $\epsilon_3 = 10^{-12}$ is used in DMRG, the result is $1.006(15)$, the most accurate. For Hamiltonian \eqref{eq:any-spin-ham-u} with $S = 1$, although the same truncation error is used, the higher-order correction of the peak height of $S^{\prime}_{vN}$ is hard to take into account, thus the result is $0.96(10)$ with a larger uncertainty, but still consistent with $c = 1$. For other cases [Hamiltonian \eqref{eq:any-spin-ham-u} with $S = 3,4$ and Hamiltonian \eqref{eq:any-spin-ham-ladder} with $S = 2,3,4,5$] that have a larger truncation error $\epsilon_2 = 10^{-11}$ in DMRG calculations, most of the results still have consistent values of $c$, but with a larger uncertainty than that for Hamiltonian \eqref{eq:any-spin-ham-u} with $S = 2$. The results for Hamiltonian \eqref{eq:any-spin-ham-u} with $S = 3, 4$ are a little off $c = 1$ even within uncertainties. These results all meet our expectation. As the finite-size effects are strong due to the logarithmic scaling, the extrapolation is sensitive to the accuracy of the data. We calculate $S^{\prime}_{vN}$ by taking a numeric differentiation, where lots of significant numbers are subtracted. Thus we need a small truncation error in DMRG to generate accurate data.

Comparing $\chi_F$ and $S^{\prime}_{vN}$, the computational procedure for the two quantities is the same. They both require calculations of the ground states at two close values of $D$ by DMRG. For the same system size, the peak position of $\chi_F$ is larger than that of $S^{\prime}_{vN}$. Systems with larger $D$ are deeper inside the gapped phase and have lower entanglement entropy, thus require cheaper computational resources in DMRG. We can compute a single point for $L = 2048, S = 1$ around the peak of $\chi_F$ within five days, while we need more than ten days to compute a single point for $L = 1024, S = 1$ around the peak of $S^{\prime}_{vN}$. Because the scalings of the peak heights of $\chi_F$ and $S^{\prime}_{vN}$ can both signal the existence of the IOG transitions or the BKT transitions, differentiating between the two IOQPTs can be faster by using $\chi_F$ than using $S^{\prime}_{vN}$. But again, there is no singularity in $\chi_F$, thus the peak position is not guaranteed to be at the IOQPT point. We can extrapolate values close to BKT points using intermediate system sizes, but there is no unbiased criteria to choose the proper range of $L$. The extrapolation from the FSS of the peak of $S^{\prime}_{vN}$ is much more accurate than that of $\chi_F$, and can be crosschecked with predictions by conformal field theory. So the derivative of the block entanglement entropy with respect to the coupling constant is in general a better universal tool to detect IOQPTs.

\subsection{Crossing points} \label{subsec:crosspoint}
\begin{figure}
  \centering
    \includegraphics[width=0.48\textwidth]{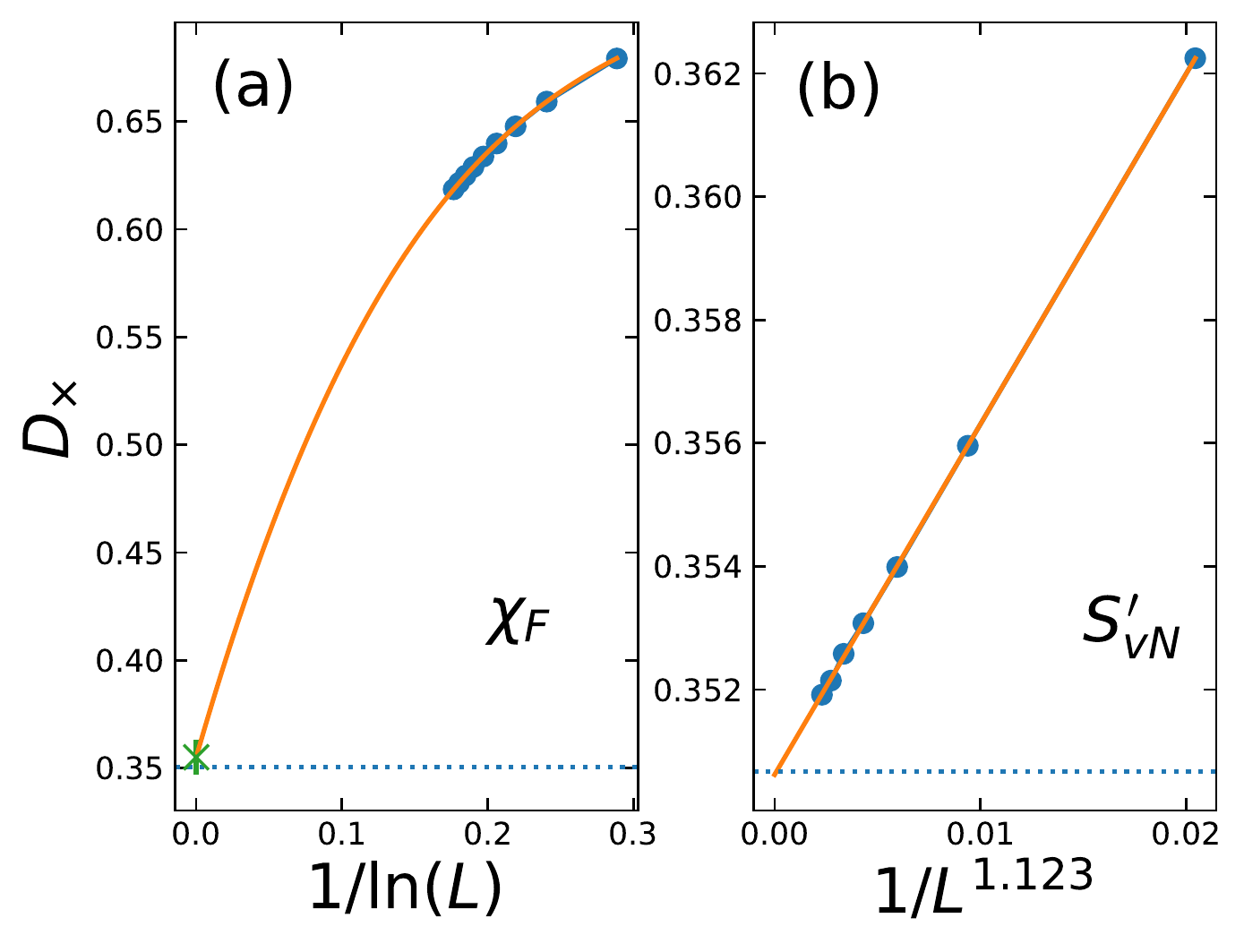}
    \caption{\label{fig:s1crossextrap}Extrapolations of the crossing points $D_{\times}$ for (a) $\chi_F$ and (b) $S^{\prime}_{vN}$. The results are for Hamiltonian \eqref{eq:any-spin-ham-u} with $S = 1$ truncation. The data in (a) are fit with a polynomial of $1/\ln(L)$, and the extrapolated $D_{\times} = 0.355(8)$. The data in (b) are fit with a pow-law of $1/L$, and the extrapolated $D_{\times} = 0.35062(6)$. The dashed line is the result from level spectroscopy $0.35066928$ \cite{PhysRevB.103.245137}.
    }
  \end{figure}

\begin{figure}
  \centering
    \includegraphics[width=0.48\textwidth]{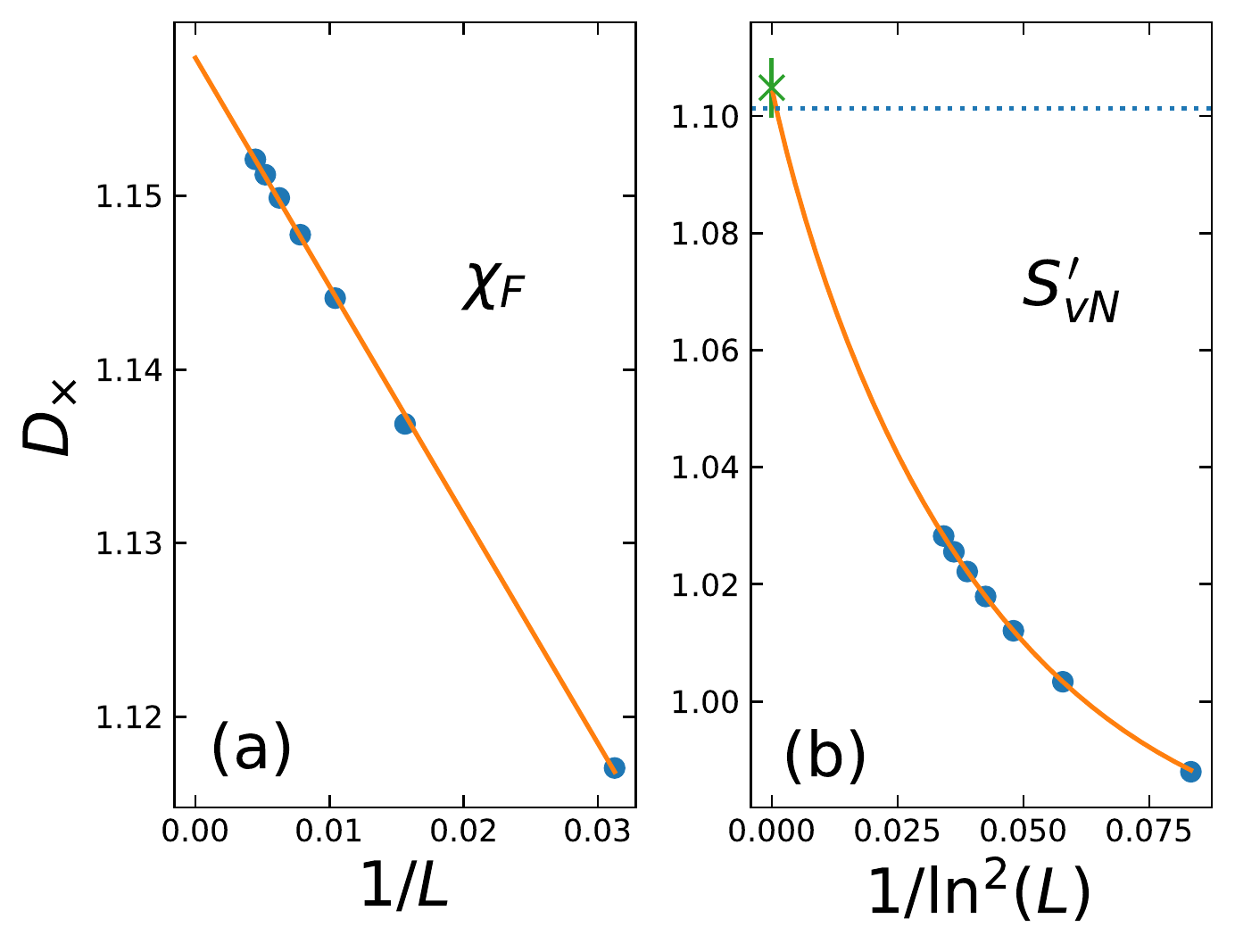}
    \caption{\label{fig:s2crossextrap}Same as Fig.~\ref{fig:s1crossextrap}, but for Hamiltonian \eqref{eq:any-spin-ham-u} with $S = 2$ truncation. The data in (a) are fit with a linear function of $1/L$ and the extrapolated $D_{\times} = 1.15794(17)$. The data in (b) are fit with Eq.~\eqref{eq:delDvslnLs2} and the extrapolated $D_{\times} = 1.105(5)$. The dashed line is the result from level spectroscopy $1.101304$ \cite{PhysRevB.103.245137}.
    }
  \end{figure}
The crossing points between different system sizes are widely seen in quantities for models near BKT transitions, such as the rescaled spin stiffness for the two-dimensional XY model \cite{Hsieh_2013}, the rescaled resistence in two-dimensional Coulomb gas \cite{PhysRevB.51.6163}, and the rescaled energy gap for quantum Hamiltonians \cite{PhysRevA.87.043606, PhysRevB.91.165136, PhysRevB.103.245137}. There also exist crossing points in both $\chi_F$ and $S^{\prime}_{vN}$ as shown in Fig.~\ref{fig:s1234fidsus} and Fig.~\ref{fig:s1234dsdy}, respectively. In Fig.~\ref{fig:s1234fidsus}, the crossing point of $\chi_F$ at $S = 1$ seems to be around $0.65$, far from the IOG point, while those at $S \ge 2$ are all close to the BKT points. In Fig.~\ref{fig:s1234dsdy}, all the crossing points of $S^{\prime}_{vN}$ are close to the phase-transition points. To locate the crossing point in the thermodynamic limit, we find the crossing point $D_{\times}$ between system sizes $L$ and $L+32$ and study the FSS. The results for $\chi_F$ and $S^{\prime}_{vN}$ at $S = 1$ are presented in Fig.~\ref{fig:s1crossextrap}. Interestingly, although the peak position of $\chi_F$ is far from the IOG point, using a polynomial function of $1/\ln(L)$ for the extrapolation, the thermodynamic position of the crossing point is extrapolated to $0.355(8)$, consistent with the location of the IOG point. For $S^{\prime}_{vN}$, we find that the crossing point is a power-law function of $1/L$, $D_{\times} \sim 1/L^{1.123(15)}$, where the power is consistent with the value of the scaling dimension in Gaussian CFT \cite{Nomura_1998}. The extrapolated position is $0.35062(6)$, which has a difference only of order $10^{-5}$ from the IOG transition point from LS. Thus, both the crossing point of $\chi_F$ and that of $S^{\prime}_{vN}$ can be used to locate the IOG point. If no high precision is needed, the calculation of the crossing point of $\chi_F$ in DMRG is much faster than that of $S^{\prime}_{vN}$ since it is at much larger $D$ (with much lower entanglement entropy) for the same system size. 

The FSS of the crossing point of $\chi_F$ for $S = 2$ is depicted in Fig.~\ref{fig:s2crossextrap}(a), where it scales linearly with $1/L$ instead of $1/\ln^2(L)$. This means that the crossing point is not associated with the renormalization group of the marginal operator around the BKT point. The extrapolated location of the crossing point is $1.15794(17)$, larger than the BKT point. In fact, the value of $D_{\times}$ for $L = 32$ is already larger than the BKT point, and it becomes larger as we increase the system size. So the peak position of $\chi_F$, which should be always larger than the crossing point, is also larger than the BKT point. Thus the extrapolated values of BKT points from peaks of $\chi_F$ for intermediate system sizes in Table \ref{tableYcSsfs} are not the true positions of the peaks. The results for $S^{\prime}_{vN}$ are shown in Fig.~\ref{fig:s2crossextrap}(b), where we fit the data with the same scaling as the peak in Eq.~\eqref{eq:delDvslnLs2} and find that the extrapolated position of the crossing point is $1.105(5)$, consistent with the value of the BKT point. So both the crossing point and the peak of $S^{\prime}_{vN}$ are located at the BKT transition points.

In a word, the FSS of the crossing point of $\chi_F$ for the IOG transition and that of $S^{\prime}_{vN}$ for BKT transitions are the same as the FSS of the peak positions of $\chi_F$ and $S^{\prime}_{vN}$, respectively, which is deduced from the divergent behavior of the correlation length. The FSS of the crossing point of $S^{\prime}_{vN}$ for the IOG transition is a power law of $1/L$ and converges to the IOG transition point, and that of $\chi_F$ for BKT transitions is linear with $1/L$ and does not characterize the BKT transitions.

\section{Conclusions} \label{sec:conclusion}
We have used a previously studied model, the truncated quantum O(2) model, to test two candidates for universal methods of detecting infinite-order quantum phase transitions (IOQPTs): the fidelity susceptibility ($\chi_F$) and the derivative of the block entanglement entropy with respect to the coupling constant ($S^{\prime}_{vN}$). Our model has an infinite-order Gaussian (IOG) transition from a gapped phase into a BKT critical line for $S = 1$ truncation, while it has a BKT transition for $S \ge 2$ truncations. The essential singularities in the correlation length are different for the two IOQPTs, which are $(D-D_c)^{-1/2}\exp\left[b_1/(D-D_c)\right]$ for $S = 1$ and $\exp(b_S/\sqrt{D-D_c})$ for $S \ge 2$. In Ref.~\cite{PhysRevB.103.245137}, we applied the level spectroscopy (LS) method and obtained accurate phase-transition points for $S = 1, 2, 3, 4, 5$ truncations. In this work, we studied the finite-size scalings (FSSs) of the peak positions and the peak heights of $\chi_F$ and $S^{\prime}_{vN}$. We elaborated how to differentiate between IOQPTs and locate the phase-transition points using the two quantities.

We showed that the peak position of $\chi_F$ in the thermodynamic limit is larger and far from the IOG point for the quantum O(2) model with $S = 1$ truncation, consistent with the observations in the fermionic Hubbard model \cite{PhysRevE.76.022101, PhysRevA.84.043601} and the $J_1$-$J_2$ Heisenberg chain \cite{PhysRevE.76.061108} that have the same type of IOG transition. Using a FSS [$A / \ln^2(BL)$] deduced from the correlation length and data for system sizes less than $512$, we extrapolated values close to the BKT points with differences of order $10^{-2}$ for $S \ge 2$ truncations. This success can be traced back to the fact that the FSS of the peak height [$\chi^*_F \sim 1/\ln(L)$] is mostly contributed by the marginal operators \cite{PhysRevB.91.014418}. We confirmed this scaling by study the FSS of $\chi^{*}_F$ for $S \ge 2$, and found that $\chi^*_F$ for the untruncated quantum O(2) model ($L \rightarrow \infty, S \rightarrow \infty$) is $0.2423(7)$. The value of $\chi^*_F$ for $S = 1$, however, has a power-law scaling with $1/L$. So the scaling of $\chi^{*}_F$ can differentiate between different types of IOQPTs. The crossing point of $\chi_F$ between different system sizes being larger than the BKT point indicates that the peak position should also be larger than the BKT point, which is also seen in the one-dimensional Bose Hubbard model with integer filling \cite{PhysRevA.87.043606} and one-dimensional SU(N) Hubbard models \cite{PhysRevA.84.043601}. We concluded that the FSS of the peak position of nonsingular $\chi_F$ for BKT transitions satisfies the scaling deduced from the correlation length only for intermediate system sizes, and the true peak positions are all shifted into the gapped phase for the IOG transition and BKT transitions. Reference \cite{PhysRevB.100.081108} also showed that there is a universal shift of the peak position of $\chi_F$ into the gapped phase for BKT transitions, consistent with our findings. 

We next investigated the FSS of $S^{\prime}_{vN}$. We found that the standard FSS of the peak position with several hundreds of sites can be used to extrapolate accurate values of both the IOG point and BKT points with differences only of order $10^{-3}$. The FSS of the crossing point of $S^{\prime}_{vN}$ between different system sizes can also predict the IOG point and BKT points accurately. The peak height of $S^{\prime}_{vN}$ diverges as $\ln^2(L)$ for the IOG transition at $S = 1$ and as $\ln^3(L)$ for BKT transitions at $S \ge 2$. Thus the scaling of the peak height can also differentiate between the two types of IOQPTs. The FSSs of the peaks are universal, thus $S^{\prime}_{vN}$ is a better tool to detect IOQPTs.

Although the peak position of $\chi_F$ is not at the IOG point or BKT points, the crossing point of $\chi_F$ at $S = 1$ is shown to be at the IOG point, which scales as $1/\ln(L)$. But it scales linearly with $1/L$ for BKT transitions at $S \ge 2$ and is larger than the BKT point inside the gapped phase. The crossing point of $S^{\prime}_{vN}$ at $S = 1$ scales as a power law of $1/L$, while it scales as $1/\ln^2(L)$ for BKT transitions. 
According to the results in other works, the crossing point of $\chi_F$ is not seen in the gapless-phase side of the spin-$1/2$ $XXZ$ chain \cite{PhysRevA.81.064301, PhysRevB.91.014418}, the extended Bose-Hubbard model \cite{PhysRevB.95.085102}, or the $Z_p$ ($p \ge 5$) clock models \cite{PhysRevB.100.094428}, but is seen in the ordered-phase side. The crossing point of $S^{\prime}_{vN}$ is also seen at the BKT point of the spin-$1/2$ $XXZ$ chain \cite{PhysRevA.81.032334}. It is interesting to investigate if the scaling behaviors of crossing points are universal in the future work.

Finally, although $\chi_F$ and $S^{\prime}_{vN}$ can be calculated accurately in one-dimensional systems with the powerful DMRG technique, many significant digits are subtracted in numerical differentiations thus computational resources are wasted. Calculating them accurately in higher-dimensional systems would require formidable work, and only exact diagonalization of small systems \cite{PhysRevB.80.094529,PhysRevB.84.174426,PhysRevB.84.125113,PhysRevA.89.043612,Nishiyama2019} are possible. We notice that there have been algorithms developed to calculate $\chi_F$ without numerical differentiation \cite{PhysRevLett.105.117203,PhysRevB.81.064418,PhysRevX.5.031007}, which makes it possible for calculations of $\chi_F$ in higher dimensions. It would also be great to develop algorithms to calculate $S^{\prime}_{vN}$ and avoid numerical differentiation.

\vskip20pt
\begin{acknowledgments}
We thank G. Ortiz, Y. Meurice, and S.-W. Tsai for helpful discussions. This work was supported by the U.S. Department of Energy (DOE) under Award Number DE-SC0019139. Computations were performed using the computer clusters and data storage resources of the HPCC, which were funded by grants from NSF (MRI-1429826) and NIH (1S10OD016290-01A1).
\end{acknowledgments}


\begin{thebibliography}{76}%
\makeatletter
\providecommand \@ifxundefined [1]{%
 \@ifx{#1\undefined}
}%
\providecommand \@ifnum [1]{%
 \ifnum #1\expandafter \@firstoftwo
 \else \expandafter \@secondoftwo
 \fi
}%
\providecommand \@ifx [1]{%
 \ifx #1\expandafter \@firstoftwo
 \else \expandafter \@secondoftwo
 \fi
}%
\providecommand \natexlab [1]{#1}%
\providecommand \enquote  [1]{``#1''}%
\providecommand \bibnamefont  [1]{#1}%
\providecommand \bibfnamefont [1]{#1}%
\providecommand \citenamefont [1]{#1}%
\providecommand \href@noop [0]{\@secondoftwo}%
\providecommand \href [0]{\begingroup \@sanitize@url \@href}%
\providecommand \@href[1]{\@@startlink{#1}\@@href}%
\providecommand \@@href[1]{\endgroup#1\@@endlink}%
\providecommand \@sanitize@url [0]{\catcode `\\12\catcode `\$12\catcode
  `\&12\catcode `\#12\catcode `\^12\catcode `\_12\catcode `\%12\relax}%
\providecommand \@@startlink[1]{}%
\providecommand \@@endlink[0]{}%
\providecommand \url  [0]{\begingroup\@sanitize@url \@url }%
\providecommand \@url [1]{\endgroup\@href {#1}{\urlprefix }}%
\providecommand \urlprefix  [0]{URL }%
\providecommand \Eprint [0]{\href }%
\providecommand \doibase [0]{http://dx.doi.org/}%
\providecommand \selectlanguage [0]{\@gobble}%
\providecommand \bibinfo  [0]{\@secondoftwo}%
\providecommand \bibfield  [0]{\@secondoftwo}%
\providecommand \translation [1]{[#1]}%
\providecommand \BibitemOpen [0]{}%
\providecommand \bibitemStop [0]{}%
\providecommand \bibitemNoStop [0]{.\EOS\space}%
\providecommand \EOS [0]{\spacefactor3000\relax}%
\providecommand \BibitemShut  [1]{\csname bibitem#1\endcsname}%
\let\auto@bib@innerbib\@empty
\bibitem [{\citenamefont {Zanardi}\ and\ \citenamefont
  {Paunkovi\ifmmode~\acute{c}\else \'{c}\fi{}}(2006)}]{PhysRevE.74.031123}%
  \BibitemOpen
  \bibfield  {author} {\bibinfo {author} {\bibfnamefont {P.}~\bibnamefont
  {Zanardi}}\ and\ \bibinfo {author} {\bibfnamefont {N.}~\bibnamefont
  {Paunkovi\ifmmode~\acute{c}\else \'{c}\fi{}}},\ }\href {\doibase
  10.1103/PhysRevE.74.031123} {\bibfield  {journal} {\bibinfo  {journal} {Phys.
  Rev. E}\ }\textbf {\bibinfo {volume} {74}},\ \bibinfo {pages} {031123}
  (\bibinfo {year} {2006})}\BibitemShut {NoStop}%
\bibitem [{\citenamefont {Chen}\ \emph {et~al.}(2007)\citenamefont {Chen},
  \citenamefont {Wang}, \citenamefont {Gu},\ and\ \citenamefont
  {Wang}}]{PhysRevE.76.061108}%
  \BibitemOpen
  \bibfield  {author} {\bibinfo {author} {\bibfnamefont {S.}~\bibnamefont
  {Chen}}, \bibinfo {author} {\bibfnamefont {L.}~\bibnamefont {Wang}}, \bibinfo
  {author} {\bibfnamefont {S.-J.}\ \bibnamefont {Gu}}, \ and\ \bibinfo {author}
  {\bibfnamefont {Y.}~\bibnamefont {Wang}},\ }\href {\doibase
  10.1103/PhysRevE.76.061108} {\bibfield  {journal} {\bibinfo  {journal} {Phys.
  Rev. E}\ }\textbf {\bibinfo {volume} {76}},\ \bibinfo {pages} {061108}
  (\bibinfo {year} {2007})}\BibitemShut {NoStop}%
\bibitem [{\citenamefont {Zanardi}\ \emph {et~al.}(2007)\citenamefont
  {Zanardi}, \citenamefont {Quan}, \citenamefont {Wang},\ and\ \citenamefont
  {Sun}}]{PhysRevA.75.032109}%
  \BibitemOpen
  \bibfield  {author} {\bibinfo {author} {\bibfnamefont {P.}~\bibnamefont
  {Zanardi}}, \bibinfo {author} {\bibfnamefont {H.~T.}\ \bibnamefont {Quan}},
  \bibinfo {author} {\bibfnamefont {X.}~\bibnamefont {Wang}}, \ and\ \bibinfo
  {author} {\bibfnamefont {C.~P.}\ \bibnamefont {Sun}},\ }\href {\doibase
  10.1103/PhysRevA.75.032109} {\bibfield  {journal} {\bibinfo  {journal} {Phys.
  Rev. A}\ }\textbf {\bibinfo {volume} {75}},\ \bibinfo {pages} {032109}
  (\bibinfo {year} {2007})}\BibitemShut {NoStop}%
\bibitem [{\citenamefont {Chen}\ \emph {et~al.}(2008)\citenamefont {Chen},
  \citenamefont {Wang}, \citenamefont {Hao},\ and\ \citenamefont
  {Wang}}]{PhysRevA.77.032111}%
  \BibitemOpen
  \bibfield  {author} {\bibinfo {author} {\bibfnamefont {S.}~\bibnamefont
  {Chen}}, \bibinfo {author} {\bibfnamefont {L.}~\bibnamefont {Wang}}, \bibinfo
  {author} {\bibfnamefont {Y.}~\bibnamefont {Hao}}, \ and\ \bibinfo {author}
  {\bibfnamefont {Y.}~\bibnamefont {Wang}},\ }\href {\doibase
  10.1103/PhysRevA.77.032111} {\bibfield  {journal} {\bibinfo  {journal} {Phys.
  Rev. A}\ }\textbf {\bibinfo {volume} {77}},\ \bibinfo {pages} {032111}
  (\bibinfo {year} {2008})}\BibitemShut {NoStop}%
\bibitem [{\citenamefont {Tzeng}\ and\ \citenamefont
  {Yang}(2008)}]{PhysRevA.77.012311}%
  \BibitemOpen
  \bibfield  {author} {\bibinfo {author} {\bibfnamefont {Y.-C.}\ \bibnamefont
  {Tzeng}}\ and\ \bibinfo {author} {\bibfnamefont {M.-F.}\ \bibnamefont
  {Yang}},\ }\href {\doibase 10.1103/PhysRevA.77.012311} {\bibfield  {journal}
  {\bibinfo  {journal} {Phys. Rev. A}\ }\textbf {\bibinfo {volume} {77}},\
  \bibinfo {pages} {012311} (\bibinfo {year} {2008})}\BibitemShut {NoStop}%
\bibitem [{\citenamefont {Tzeng}\ \emph {et~al.}(2008)\citenamefont {Tzeng},
  \citenamefont {Hung}, \citenamefont {Chen},\ and\ \citenamefont
  {Yang}}]{PhysRevA.77.062321}%
  \BibitemOpen
  \bibfield  {author} {\bibinfo {author} {\bibfnamefont {Y.-C.}\ \bibnamefont
  {Tzeng}}, \bibinfo {author} {\bibfnamefont {H.-H.}\ \bibnamefont {Hung}},
  \bibinfo {author} {\bibfnamefont {Y.-C.}\ \bibnamefont {Chen}}, \ and\
  \bibinfo {author} {\bibfnamefont {M.-F.}\ \bibnamefont {Yang}},\ }\href
  {\doibase 10.1103/PhysRevA.77.062321} {\bibfield  {journal} {\bibinfo
  {journal} {Phys. Rev. A}\ }\textbf {\bibinfo {volume} {77}},\ \bibinfo
  {pages} {062321} (\bibinfo {year} {2008})}\BibitemShut {NoStop}%
\bibitem [{\citenamefont {Dai}\ \emph {et~al.}(2010)\citenamefont {Dai},
  \citenamefont {Hu}, \citenamefont {Zhao},\ and\ \citenamefont
  {Zhou}}]{Dai_2010}%
  \BibitemOpen
  \bibfield  {author} {\bibinfo {author} {\bibfnamefont {Y.-W.}\ \bibnamefont
  {Dai}}, \bibinfo {author} {\bibfnamefont {B.-Q.}\ \bibnamefont {Hu}},
  \bibinfo {author} {\bibfnamefont {J.-H.}\ \bibnamefont {Zhao}}, \ and\
  \bibinfo {author} {\bibfnamefont {H.-Q.}\ \bibnamefont {Zhou}},\ }\href
  {\doibase 10.1088/1751-8113/43/37/372001} {\bibfield  {journal} {\bibinfo
  {journal} {Journal of Physics A: Mathematical and Theoretical}\ }\textbf
  {\bibinfo {volume} {43}},\ \bibinfo {pages} {372001} (\bibinfo {year}
  {2010})}\BibitemShut {NoStop}%
\bibitem [{\citenamefont {Wang}\ \emph
  {et~al.}(2011{\natexlab{a}})\citenamefont {Wang}, \citenamefont {Chen},
  \citenamefont {Li},\ and\ \citenamefont {Zhou}}]{Wang_2011(1)}%
  \BibitemOpen
  \bibfield  {author} {\bibinfo {author} {\bibfnamefont {H.-L.}\ \bibnamefont
  {Wang}}, \bibinfo {author} {\bibfnamefont {A.-M.}\ \bibnamefont {Chen}},
  \bibinfo {author} {\bibfnamefont {B.}~\bibnamefont {Li}}, \ and\ \bibinfo
  {author} {\bibfnamefont {H.-Q.}\ \bibnamefont {Zhou}},\ }\href {\doibase
  10.1088/1751-8113/45/1/015306} {\bibfield  {journal} {\bibinfo  {journal}
  {Journal of Physics A: Mathematical and Theoretical}\ }\textbf {\bibinfo
  {volume} {45}},\ \bibinfo {pages} {015306} (\bibinfo {year}
  {2011}{\natexlab{a}})}\BibitemShut {NoStop}%
\bibitem [{\citenamefont {Wang}\ \emph
  {et~al.}(2011{\natexlab{b}})\citenamefont {Wang}, \citenamefont {Zhao},
  \citenamefont {Li},\ and\ \citenamefont {Zhou}}]{Wang_2011(2)}%
  \BibitemOpen
  \bibfield  {author} {\bibinfo {author} {\bibfnamefont {H.-L.}\ \bibnamefont
  {Wang}}, \bibinfo {author} {\bibfnamefont {J.-H.}\ \bibnamefont {Zhao}},
  \bibinfo {author} {\bibfnamefont {B.}~\bibnamefont {Li}}, \ and\ \bibinfo
  {author} {\bibfnamefont {H.-Q.}\ \bibnamefont {Zhou}},\ }\href {\doibase
  10.1088/1742-5468/2011/10/l10001} {\bibfield  {journal} {\bibinfo  {journal}
  {Journal of Statistical Mechanics: Theory and Experiment}\ }\textbf {\bibinfo
  {volume} {2011}},\ \bibinfo {pages} {L10001} (\bibinfo {year}
  {2011}{\natexlab{b}})}\BibitemShut {NoStop}%
\bibitem [{\citenamefont {Zhao}\ \emph {et~al.}(2010)\citenamefont {Zhao},
  \citenamefont {Wang}, \citenamefont {Li},\ and\ \citenamefont
  {Zhou}}]{PhysRevE.82.061127}%
  \BibitemOpen
  \bibfield  {author} {\bibinfo {author} {\bibfnamefont {J.-H.}\ \bibnamefont
  {Zhao}}, \bibinfo {author} {\bibfnamefont {H.-L.}\ \bibnamefont {Wang}},
  \bibinfo {author} {\bibfnamefont {B.}~\bibnamefont {Li}}, \ and\ \bibinfo
  {author} {\bibfnamefont {H.-Q.}\ \bibnamefont {Zhou}},\ }\href {\doibase
  10.1103/PhysRevE.82.061127} {\bibfield  {journal} {\bibinfo  {journal} {Phys.
  Rev. E}\ }\textbf {\bibinfo {volume} {82}},\ \bibinfo {pages} {061127}
  (\bibinfo {year} {2010})}\BibitemShut {NoStop}%
\bibitem [{\citenamefont {Thesberg}\ and\ \citenamefont
  {S\o{}rensen}(2011)}]{PhysRevB.84.224435}%
  \BibitemOpen
  \bibfield  {author} {\bibinfo {author} {\bibfnamefont {M.}~\bibnamefont
  {Thesberg}}\ and\ \bibinfo {author} {\bibfnamefont {E.~S.}\ \bibnamefont
  {S\o{}rensen}},\ }\href {\doibase 10.1103/PhysRevB.84.224435} {\bibfield
  {journal} {\bibinfo  {journal} {Phys. Rev. B}\ }\textbf {\bibinfo {volume}
  {84}},\ \bibinfo {pages} {224435} (\bibinfo {year} {2011})}\BibitemShut
  {NoStop}%
\bibitem [{\citenamefont {\L{}acki}\ \emph {et~al.}(2014)\citenamefont
  {\L{}acki}, \citenamefont {Damski},\ and\ \citenamefont
  {Zakrzewski}}]{PhysRevA.89.033625}%
  \BibitemOpen
  \bibfield  {author} {\bibinfo {author} {\bibfnamefont {M.}~\bibnamefont
  {\L{}acki}}, \bibinfo {author} {\bibfnamefont {B.}~\bibnamefont {Damski}}, \
  and\ \bibinfo {author} {\bibfnamefont {J.}~\bibnamefont {Zakrzewski}},\
  }\href {\doibase 10.1103/PhysRevA.89.033625} {\bibfield  {journal} {\bibinfo
  {journal} {Phys. Rev. A}\ }\textbf {\bibinfo {volume} {89}},\ \bibinfo
  {pages} {033625} (\bibinfo {year} {2014})}\BibitemShut {NoStop}%
\bibitem [{\citenamefont {Wei}\ and\ \citenamefont
  {Lv}(2018)}]{PhysRevA.97.013845}%
  \BibitemOpen
  \bibfield  {author} {\bibinfo {author} {\bibfnamefont {B.-B.}\ \bibnamefont
  {Wei}}\ and\ \bibinfo {author} {\bibfnamefont {X.-C.}\ \bibnamefont {Lv}},\
  }\href {\doibase 10.1103/PhysRevA.97.013845} {\bibfield  {journal} {\bibinfo
  {journal} {Phys. Rev. A}\ }\textbf {\bibinfo {volume} {97}},\ \bibinfo
  {pages} {013845} (\bibinfo {year} {2018})}\BibitemShut {NoStop}%
\bibitem [{\citenamefont {\DJ{}uri\ifmmode~\acute{c}\else \'{c}\fi{}}\ \emph
  {et~al.}(2017)\citenamefont {\DJ{}uri\ifmmode~\acute{c}\else \'{c}\fi{}},
  \citenamefont {Biedro\ifmmode~\acute{n}\else \'{n}\fi{}},\ and\ \citenamefont
  {Zakrzewski}}]{PhysRevB.95.085102}%
  \BibitemOpen
  \bibfield  {author} {\bibinfo {author} {\bibfnamefont {T.}~\bibnamefont
  {\DJ{}uri\ifmmode~\acute{c}\else \'{c}\fi{}}}, \bibinfo {author}
  {\bibfnamefont {K.}~\bibnamefont {Biedro\ifmmode~\acute{n}\else \'{n}\fi{}}},
  \ and\ \bibinfo {author} {\bibfnamefont {J.}~\bibnamefont {Zakrzewski}},\
  }\href {\doibase 10.1103/PhysRevB.95.085102} {\bibfield  {journal} {\bibinfo
  {journal} {Phys. Rev. B}\ }\textbf {\bibinfo {volume} {95}},\ \bibinfo
  {pages} {085102} (\bibinfo {year} {2017})}\BibitemShut {NoStop}%
\bibitem [{\citenamefont {Zhu}\ \emph {et~al.}(2018)\citenamefont {Zhu},
  \citenamefont {Sun}, \citenamefont {You},\ and\ \citenamefont
  {Shi}}]{PhysRevA.98.023607}%
  \BibitemOpen
  \bibfield  {author} {\bibinfo {author} {\bibfnamefont {Z.}~\bibnamefont
  {Zhu}}, \bibinfo {author} {\bibfnamefont {G.}~\bibnamefont {Sun}}, \bibinfo
  {author} {\bibfnamefont {W.-L.}\ \bibnamefont {You}}, \ and\ \bibinfo
  {author} {\bibfnamefont {D.-N.}\ \bibnamefont {Shi}},\ }\href {\doibase
  10.1103/PhysRevA.98.023607} {\bibfield  {journal} {\bibinfo  {journal} {Phys.
  Rev. A}\ }\textbf {\bibinfo {volume} {98}},\ \bibinfo {pages} {023607}
  (\bibinfo {year} {2018})}\BibitemShut {NoStop}%
\bibitem [{\citenamefont {Luo}\ \emph {et~al.}(2018)\citenamefont {Luo},
  \citenamefont {Zhao},\ and\ \citenamefont {Wang}}]{PhysRevE.98.022106}%
  \BibitemOpen
  \bibfield  {author} {\bibinfo {author} {\bibfnamefont {Q.}~\bibnamefont
  {Luo}}, \bibinfo {author} {\bibfnamefont {J.}~\bibnamefont {Zhao}}, \ and\
  \bibinfo {author} {\bibfnamefont {X.}~\bibnamefont {Wang}},\ }\href {\doibase
  10.1103/PhysRevE.98.022106} {\bibfield  {journal} {\bibinfo  {journal} {Phys.
  Rev. E}\ }\textbf {\bibinfo {volume} {98}},\ \bibinfo {pages} {022106}
  (\bibinfo {year} {2018})}\BibitemShut {NoStop}%
\bibitem [{\citenamefont {Sun}\ \emph {et~al.}(2019)\citenamefont {Sun},
  \citenamefont {Vekua}, \citenamefont {Cobanera},\ and\ \citenamefont
  {Ortiz}}]{PhysRevB.100.094428}%
  \BibitemOpen
  \bibfield  {author} {\bibinfo {author} {\bibfnamefont {G.}~\bibnamefont
  {Sun}}, \bibinfo {author} {\bibfnamefont {T.}~\bibnamefont {Vekua}}, \bibinfo
  {author} {\bibfnamefont {E.}~\bibnamefont {Cobanera}}, \ and\ \bibinfo
  {author} {\bibfnamefont {G.}~\bibnamefont {Ortiz}},\ }\href {\doibase
  10.1103/PhysRevB.100.094428} {\bibfield  {journal} {\bibinfo  {journal}
  {Phys. Rev. B}\ }\textbf {\bibinfo {volume} {100}},\ \bibinfo {pages}
  {094428} (\bibinfo {year} {2019})}\BibitemShut {NoStop}%
\bibitem [{\citenamefont {Osterloh}\ \emph {et~al.}(2002)\citenamefont
  {Osterloh}, \citenamefont {Amico}, \citenamefont {Falci},\ and\ \citenamefont
  {Fazio}}]{Osterloh2002}%
  \BibitemOpen
  \bibfield  {author} {\bibinfo {author} {\bibfnamefont {A.}~\bibnamefont
  {Osterloh}}, \bibinfo {author} {\bibfnamefont {L.}~\bibnamefont {Amico}},
  \bibinfo {author} {\bibfnamefont {G.}~\bibnamefont {Falci}}, \ and\ \bibinfo
  {author} {\bibfnamefont {R.}~\bibnamefont {Fazio}},\ }\href {\doibase
  10.1038/416608a} {\bibfield  {journal} {\bibinfo  {journal} {Nature}\
  }\textbf {\bibinfo {volume} {416}},\ \bibinfo {pages} {608} (\bibinfo {year}
  {2002})}\BibitemShut {NoStop}%
\bibitem [{\citenamefont {Osborne}\ and\ \citenamefont
  {Nielsen}(2002)}]{PhysRevA.66.032110}%
  \BibitemOpen
  \bibfield  {author} {\bibinfo {author} {\bibfnamefont {T.~J.}\ \bibnamefont
  {Osborne}}\ and\ \bibinfo {author} {\bibfnamefont {M.~A.}\ \bibnamefont
  {Nielsen}},\ }\href {\doibase 10.1103/PhysRevA.66.032110} {\bibfield
  {journal} {\bibinfo  {journal} {Phys. Rev. A}\ }\textbf {\bibinfo {volume}
  {66}},\ \bibinfo {pages} {032110} (\bibinfo {year} {2002})}\BibitemShut
  {NoStop}%
\bibitem [{\citenamefont {Gu}\ \emph {et~al.}(2004)\citenamefont {Gu},
  \citenamefont {Deng}, \citenamefont {Li},\ and\ \citenamefont
  {Lin}}]{PhysRevLett.93.086402}%
  \BibitemOpen
  \bibfield  {author} {\bibinfo {author} {\bibfnamefont {S.-J.}\ \bibnamefont
  {Gu}}, \bibinfo {author} {\bibfnamefont {S.-S.}\ \bibnamefont {Deng}},
  \bibinfo {author} {\bibfnamefont {Y.-Q.}\ \bibnamefont {Li}}, \ and\ \bibinfo
  {author} {\bibfnamefont {H.-Q.}\ \bibnamefont {Lin}},\ }\href {\doibase
  10.1103/PhysRevLett.93.086402} {\bibfield  {journal} {\bibinfo  {journal}
  {Phys. Rev. Lett.}\ }\textbf {\bibinfo {volume} {93}},\ \bibinfo {pages}
  {086402} (\bibinfo {year} {2004})}\BibitemShut {NoStop}%
\bibitem [{\citenamefont {Wu}\ \emph {et~al.}(2004)\citenamefont {Wu},
  \citenamefont {Sarandy},\ and\ \citenamefont
  {Lidar}}]{PhysRevLett.93.250404}%
  \BibitemOpen
  \bibfield  {author} {\bibinfo {author} {\bibfnamefont {L.-A.}\ \bibnamefont
  {Wu}}, \bibinfo {author} {\bibfnamefont {M.~S.}\ \bibnamefont {Sarandy}}, \
  and\ \bibinfo {author} {\bibfnamefont {D.~A.}\ \bibnamefont {Lidar}},\ }\href
  {\doibase 10.1103/PhysRevLett.93.250404} {\bibfield  {journal} {\bibinfo
  {journal} {Phys. Rev. Lett.}\ }\textbf {\bibinfo {volume} {93}},\ \bibinfo
  {pages} {250404} (\bibinfo {year} {2004})}\BibitemShut {NoStop}%
\bibitem [{\citenamefont {Larsson}\ and\ \citenamefont
  {Johannesson}(2005)}]{PhysRevLett.95.196406}%
  \BibitemOpen
  \bibfield  {author} {\bibinfo {author} {\bibfnamefont {D.}~\bibnamefont
  {Larsson}}\ and\ \bibinfo {author} {\bibfnamefont {H.}~\bibnamefont
  {Johannesson}},\ }\href {\doibase 10.1103/PhysRevLett.95.196406} {\bibfield
  {journal} {\bibinfo  {journal} {Phys. Rev. Lett.}\ }\textbf {\bibinfo
  {volume} {95}},\ \bibinfo {pages} {196406} (\bibinfo {year}
  {2005})}\BibitemShut {NoStop}%
\bibitem [{\citenamefont {Larsson}\ and\ \citenamefont
  {Johannesson}(2006)}]{PhysRevA.73.042320}%
  \BibitemOpen
  \bibfield  {author} {\bibinfo {author} {\bibfnamefont {D.}~\bibnamefont
  {Larsson}}\ and\ \bibinfo {author} {\bibfnamefont {H.}~\bibnamefont
  {Johannesson}},\ }\href {\doibase 10.1103/PhysRevA.73.042320} {\bibfield
  {journal} {\bibinfo  {journal} {Phys. Rev. A}\ }\textbf {\bibinfo {volume}
  {73}},\ \bibinfo {pages} {042320} (\bibinfo {year} {2006})}\BibitemShut
  {NoStop}%
\bibitem [{\citenamefont {Legeza}\ and\ \citenamefont
  {S\'olyom}(2006)}]{PhysRevLett.96.116401}%
  \BibitemOpen
  \bibfield  {author} {\bibinfo {author} {\bibfnamefont {O.}~\bibnamefont
  {Legeza}}\ and\ \bibinfo {author} {\bibfnamefont {J.}~\bibnamefont
  {S\'olyom}},\ }\href {\doibase 10.1103/PhysRevLett.96.116401} {\bibfield
  {journal} {\bibinfo  {journal} {Phys. Rev. Lett.}\ }\textbf {\bibinfo
  {volume} {96}},\ \bibinfo {pages} {116401} (\bibinfo {year}
  {2006})}\BibitemShut {NoStop}%
\bibitem [{\citenamefont {Campos~Venuti}\ \emph {et~al.}(2006)\citenamefont
  {Campos~Venuti}, \citenamefont {Degli Esposti~Boschi}, \citenamefont
  {Roncaglia},\ and\ \citenamefont {Scaramucci}}]{PhysRevA.73.010303}%
  \BibitemOpen
  \bibfield  {author} {\bibinfo {author} {\bibfnamefont {L.}~\bibnamefont
  {Campos~Venuti}}, \bibinfo {author} {\bibfnamefont {C.}~\bibnamefont {Degli
  Esposti~Boschi}}, \bibinfo {author} {\bibfnamefont {M.}~\bibnamefont
  {Roncaglia}}, \ and\ \bibinfo {author} {\bibfnamefont {A.}~\bibnamefont
  {Scaramucci}},\ }\href {\doibase 10.1103/PhysRevA.73.010303} {\bibfield
  {journal} {\bibinfo  {journal} {Phys. Rev. A}\ }\textbf {\bibinfo {volume}
  {73}},\ \bibinfo {pages} {010303} (\bibinfo {year} {2006})}\BibitemShut
  {NoStop}%
\bibitem [{\citenamefont {Wu}\ \emph {et~al.}(2006)\citenamefont {Wu},
  \citenamefont {Sarandy}, \citenamefont {Lidar},\ and\ \citenamefont
  {Sham}}]{PhysRevA.74.052335}%
  \BibitemOpen
  \bibfield  {author} {\bibinfo {author} {\bibfnamefont {L.-A.}\ \bibnamefont
  {Wu}}, \bibinfo {author} {\bibfnamefont {M.~S.}\ \bibnamefont {Sarandy}},
  \bibinfo {author} {\bibfnamefont {D.~A.}\ \bibnamefont {Lidar}}, \ and\
  \bibinfo {author} {\bibfnamefont {L.~J.}\ \bibnamefont {Sham}},\ }\href
  {\doibase 10.1103/PhysRevA.74.052335} {\bibfield  {journal} {\bibinfo
  {journal} {Phys. Rev. A}\ }\textbf {\bibinfo {volume} {74}},\ \bibinfo
  {pages} {052335} (\bibinfo {year} {2006})}\BibitemShut {NoStop}%
\bibitem [{\citenamefont {Rulli}\ and\ \citenamefont
  {Sarandy}(2010)}]{PhysRevA.81.032334}%
  \BibitemOpen
  \bibfield  {author} {\bibinfo {author} {\bibfnamefont {C.~C.}\ \bibnamefont
  {Rulli}}\ and\ \bibinfo {author} {\bibfnamefont {M.~S.}\ \bibnamefont
  {Sarandy}},\ }\href {\doibase 10.1103/PhysRevA.81.032334} {\bibfield
  {journal} {\bibinfo  {journal} {Phys. Rev. A}\ }\textbf {\bibinfo {volume}
  {81}},\ \bibinfo {pages} {032334} (\bibinfo {year} {2010})}\BibitemShut
  {NoStop}%
\bibitem [{\citenamefont {Nishimoto}(2011)}]{PhysRevB.84.195108}%
  \BibitemOpen
  \bibfield  {author} {\bibinfo {author} {\bibfnamefont {S.}~\bibnamefont
  {Nishimoto}},\ }\href {\doibase 10.1103/PhysRevB.84.195108} {\bibfield
  {journal} {\bibinfo  {journal} {Phys. Rev. B}\ }\textbf {\bibinfo {volume}
  {84}},\ \bibinfo {pages} {195108} (\bibinfo {year} {2011})}\BibitemShut
  {NoStop}%
\bibitem [{\citenamefont {Li}\ \emph {et~al.}(2016)\citenamefont {Li},
  \citenamefont {Zhu},\ and\ \citenamefont {Yuan}}]{LI20161066}%
  \BibitemOpen
  \bibfield  {author} {\bibinfo {author} {\bibfnamefont {Y.-C.}\ \bibnamefont
  {Li}}, \bibinfo {author} {\bibfnamefont {Y.-H.}\ \bibnamefont {Zhu}}, \ and\
  \bibinfo {author} {\bibfnamefont {Z.-G.}\ \bibnamefont {Yuan}},\ }\href
  {\doibase https://doi.org/10.1016/j.physleta.2016.01.004} {\bibfield
  {journal} {\bibinfo  {journal} {Physics Letters A}\ }\textbf {\bibinfo
  {volume} {380}},\ \bibinfo {pages} {1066} (\bibinfo {year}
  {2016})}\BibitemShut {NoStop}%
\bibitem [{\citenamefont {Liu}\ and\ \citenamefont
  {Bhatt}(2016)}]{PhysRevLett.117.206801}%
  \BibitemOpen
  \bibfield  {author} {\bibinfo {author} {\bibfnamefont {Z.}~\bibnamefont
  {Liu}}\ and\ \bibinfo {author} {\bibfnamefont {R.~N.}\ \bibnamefont
  {Bhatt}},\ }\href {\doibase 10.1103/PhysRevLett.117.206801} {\bibfield
  {journal} {\bibinfo  {journal} {Phys. Rev. Lett.}\ }\textbf {\bibinfo
  {volume} {117}},\ \bibinfo {pages} {206801} (\bibinfo {year}
  {2016})}\BibitemShut {NoStop}%
\bibitem [{\citenamefont {Valdez}\ \emph {et~al.}(2017)\citenamefont {Valdez},
  \citenamefont {Jaschke}, \citenamefont {Vargas},\ and\ \citenamefont
  {Carr}}]{PhysRevLett.119.225301}%
  \BibitemOpen
  \bibfield  {author} {\bibinfo {author} {\bibfnamefont {M.~A.}\ \bibnamefont
  {Valdez}}, \bibinfo {author} {\bibfnamefont {D.}~\bibnamefont {Jaschke}},
  \bibinfo {author} {\bibfnamefont {D.~L.}\ \bibnamefont {Vargas}}, \ and\
  \bibinfo {author} {\bibfnamefont {L.~D.}\ \bibnamefont {Carr}},\ }\href
  {\doibase 10.1103/PhysRevLett.119.225301} {\bibfield  {journal} {\bibinfo
  {journal} {Phys. Rev. Lett.}\ }\textbf {\bibinfo {volume} {119}},\ \bibinfo
  {pages} {225301} (\bibinfo {year} {2017})}\BibitemShut {NoStop}%
\bibitem [{\citenamefont {Spalding}\ \emph {et~al.}(2019)\citenamefont
  {Spalding}, \citenamefont {Tsai},\ and\ \citenamefont
  {Campbell}}]{PhysRevB.99.195445}%
  \BibitemOpen
  \bibfield  {author} {\bibinfo {author} {\bibfnamefont {J.}~\bibnamefont
  {Spalding}}, \bibinfo {author} {\bibfnamefont {S.-W.}\ \bibnamefont {Tsai}},
  \ and\ \bibinfo {author} {\bibfnamefont {D.~K.}\ \bibnamefont {Campbell}},\
  }\href {\doibase 10.1103/PhysRevB.99.195445} {\bibfield  {journal} {\bibinfo
  {journal} {Phys. Rev. B}\ }\textbf {\bibinfo {volume} {99}},\ \bibinfo
  {pages} {195445} (\bibinfo {year} {2019})}\BibitemShut {NoStop}%
\bibitem [{\citenamefont {You}\ \emph {et~al.}(2007)\citenamefont {You},
  \citenamefont {Li},\ and\ \citenamefont {Gu}}]{PhysRevE.76.022101}%
  \BibitemOpen
  \bibfield  {author} {\bibinfo {author} {\bibfnamefont {W.-L.}\ \bibnamefont
  {You}}, \bibinfo {author} {\bibfnamefont {Y.-W.}\ \bibnamefont {Li}}, \ and\
  \bibinfo {author} {\bibfnamefont {S.-J.}\ \bibnamefont {Gu}},\ }\href
  {\doibase 10.1103/PhysRevE.76.022101} {\bibfield  {journal} {\bibinfo
  {journal} {Phys. Rev. E}\ }\textbf {\bibinfo {volume} {76}},\ \bibinfo
  {pages} {022101} (\bibinfo {year} {2007})}\BibitemShut {NoStop}%
\bibitem [{\citenamefont {Sun}\ \emph {et~al.}(2015)\citenamefont {Sun},
  \citenamefont {Kolezhuk},\ and\ \citenamefont {Vekua}}]{PhysRevB.91.014418}%
  \BibitemOpen
  \bibfield  {author} {\bibinfo {author} {\bibfnamefont {G.}~\bibnamefont
  {Sun}}, \bibinfo {author} {\bibfnamefont {A.~K.}\ \bibnamefont {Kolezhuk}}, \
  and\ \bibinfo {author} {\bibfnamefont {T.}~\bibnamefont {Vekua}},\ }\href
  {\doibase 10.1103/PhysRevB.91.014418} {\bibfield  {journal} {\bibinfo
  {journal} {Phys. Rev. B}\ }\textbf {\bibinfo {volume} {91}},\ \bibinfo
  {pages} {014418} (\bibinfo {year} {2015})}\BibitemShut {NoStop}%
\bibitem [{\citenamefont {Ding}\ and\ \citenamefont
  {Makivi\ifmmode~\acute{c}\else \'{c}\fi{}}(1990)}]{PhysRevB.42.6827}%
  \BibitemOpen
  \bibfield  {author} {\bibinfo {author} {\bibfnamefont {H.-Q.}\ \bibnamefont
  {Ding}}\ and\ \bibinfo {author} {\bibfnamefont {M.~S.}\ \bibnamefont
  {Makivi\ifmmode~\acute{c}\else \'{c}\fi{}}},\ }\href {\doibase
  10.1103/PhysRevB.42.6827} {\bibfield  {journal} {\bibinfo  {journal} {Phys.
  Rev. B}\ }\textbf {\bibinfo {volume} {42}},\ \bibinfo {pages} {6827}
  (\bibinfo {year} {1990})}\BibitemShut {NoStop}%
\bibitem [{\citenamefont {Li}\ \emph {et~al.}(2020)\citenamefont {Li},
  \citenamefont {Yang}, \citenamefont {Xie}, \citenamefont {Tu}, \citenamefont
  {Liao},\ and\ \citenamefont {Xiang}}]{PhysRevE.101.060105}%
  \BibitemOpen
  \bibfield  {author} {\bibinfo {author} {\bibfnamefont {Z.-Q.}\ \bibnamefont
  {Li}}, \bibinfo {author} {\bibfnamefont {L.-P.}\ \bibnamefont {Yang}},
  \bibinfo {author} {\bibfnamefont {Z.~Y.}\ \bibnamefont {Xie}}, \bibinfo
  {author} {\bibfnamefont {H.-H.}\ \bibnamefont {Tu}}, \bibinfo {author}
  {\bibfnamefont {H.-J.}\ \bibnamefont {Liao}}, \ and\ \bibinfo {author}
  {\bibfnamefont {T.}~\bibnamefont {Xiang}},\ }\href {\doibase
  10.1103/PhysRevE.101.060105} {\bibfield  {journal} {\bibinfo  {journal}
  {Phys. Rev. E}\ }\textbf {\bibinfo {volume} {101}},\ \bibinfo {pages}
  {060105} (\bibinfo {year} {2020})}\BibitemShut {NoStop}%
\bibitem [{\citenamefont {Calabrese}\ \emph {et~al.}(2010)\citenamefont
  {Calabrese}, \citenamefont {Campostrini}, \citenamefont {Essler},\ and\
  \citenamefont {Nienhuis}}]{PhysRevLett.104.095701}%
  \BibitemOpen
  \bibfield  {author} {\bibinfo {author} {\bibfnamefont {P.}~\bibnamefont
  {Calabrese}}, \bibinfo {author} {\bibfnamefont {M.}~\bibnamefont
  {Campostrini}}, \bibinfo {author} {\bibfnamefont {F.}~\bibnamefont {Essler}},
  \ and\ \bibinfo {author} {\bibfnamefont {B.}~\bibnamefont {Nienhuis}},\
  }\href {\doibase 10.1103/PhysRevLett.104.095701} {\bibfield  {journal}
  {\bibinfo  {journal} {Phys. Rev. Lett.}\ }\textbf {\bibinfo {volume} {104}},\
  \bibinfo {pages} {095701} (\bibinfo {year} {2010})}\BibitemShut {NoStop}%
\bibitem [{\citenamefont {Fagotti}\ and\ \citenamefont
  {Calabrese}(2011)}]{Fagotti_2011}%
  \BibitemOpen
  \bibfield  {author} {\bibinfo {author} {\bibfnamefont {M.}~\bibnamefont
  {Fagotti}}\ and\ \bibinfo {author} {\bibfnamefont {P.}~\bibnamefont
  {Calabrese}},\ }\href {\doibase 10.1088/1742-5468/2011/01/p01017} {\bibfield
  {journal} {\bibinfo  {journal} {Journal of Statistical Mechanics: Theory and
  Experiment}\ }\textbf {\bibinfo {volume} {2011}},\ \bibinfo {pages} {P01017}
  (\bibinfo {year} {2011})}\BibitemShut {NoStop}%
\bibitem [{\citenamefont {Xavier}\ and\ \citenamefont
  {Alcaraz}(2011)}]{PhysRevB.83.214425}%
  \BibitemOpen
  \bibfield  {author} {\bibinfo {author} {\bibfnamefont {J.~C.}\ \bibnamefont
  {Xavier}}\ and\ \bibinfo {author} {\bibfnamefont {F.~C.}\ \bibnamefont
  {Alcaraz}},\ }\href {\doibase 10.1103/PhysRevB.83.214425} {\bibfield
  {journal} {\bibinfo  {journal} {Phys. Rev. B}\ }\textbf {\bibinfo {volume}
  {83}},\ \bibinfo {pages} {214425} (\bibinfo {year} {2011})}\BibitemShut
  {NoStop}%
\bibitem [{\citenamefont {Chen}\ \emph {et~al.}(2003)\citenamefont {Chen},
  \citenamefont {Hida},\ and\ \citenamefont {Sanctuary}}]{PhysRevB.67.104401}%
  \BibitemOpen
  \bibfield  {author} {\bibinfo {author} {\bibfnamefont {W.}~\bibnamefont
  {Chen}}, \bibinfo {author} {\bibfnamefont {K.}~\bibnamefont {Hida}}, \ and\
  \bibinfo {author} {\bibfnamefont {B.~C.}\ \bibnamefont {Sanctuary}},\ }\href
  {\doibase 10.1103/PhysRevB.67.104401} {\bibfield  {journal} {\bibinfo
  {journal} {Phys. Rev. B}\ }\textbf {\bibinfo {volume} {67}},\ \bibinfo
  {pages} {104401} (\bibinfo {year} {2003})}\BibitemShut {NoStop}%
\bibitem [{\citenamefont {Zhang}\ \emph {et~al.}(2021)\citenamefont {Zhang},
  \citenamefont {Meurice},\ and\ \citenamefont {Tsai}}]{PhysRevB.103.245137}%
  \BibitemOpen
  \bibfield  {author} {\bibinfo {author} {\bibfnamefont {J.}~\bibnamefont
  {Zhang}}, \bibinfo {author} {\bibfnamefont {Y.}~\bibnamefont {Meurice}}, \
  and\ \bibinfo {author} {\bibfnamefont {S.-W.}\ \bibnamefont {Tsai}},\ }\href
  {\doibase 10.1103/PhysRevB.103.245137} {\bibfield  {journal} {\bibinfo
  {journal} {Phys. Rev. B}\ }\textbf {\bibinfo {volume} {103}},\ \bibinfo
  {pages} {245137} (\bibinfo {year} {2021})}\BibitemShut {NoStop}%
\bibitem [{\citenamefont {Ovchinnikov}(1970)}]{ovchinnikov1970excitation}%
  \BibitemOpen
  \bibfield  {author} {\bibinfo {author} {\bibfnamefont {A.}~\bibnamefont
  {Ovchinnikov}},\ }\href@noop {} {\bibfield  {journal} {\bibinfo  {journal}
  {Sov. Phys. JETP}\ }\textbf {\bibinfo {volume} {30}},\ \bibinfo {pages}
  {1160} (\bibinfo {year} {1970})}\BibitemShut {NoStop}%
\bibitem [{\citenamefont {Nakamura}(2000)}]{PhysRevB.61.16377}%
  \BibitemOpen
  \bibfield  {author} {\bibinfo {author} {\bibfnamefont {M.}~\bibnamefont
  {Nakamura}},\ }\href {\doibase 10.1103/PhysRevB.61.16377} {\bibfield
  {journal} {\bibinfo  {journal} {Phys. Rev. B}\ }\textbf {\bibinfo {volume}
  {61}},\ \bibinfo {pages} {16377} (\bibinfo {year} {2000})}\BibitemShut
  {NoStop}%
\bibitem [{\citenamefont {Haldane}(1982)}]{PhysRevB.25.4925}%
  \BibitemOpen
  \bibfield  {author} {\bibinfo {author} {\bibfnamefont {F.~D.~M.}\
  \bibnamefont {Haldane}},\ }\href {\doibase 10.1103/PhysRevB.25.4925}
  {\bibfield  {journal} {\bibinfo  {journal} {Phys. Rev. B}\ }\textbf {\bibinfo
  {volume} {25}},\ \bibinfo {pages} {4925} (\bibinfo {year}
  {1982})}\BibitemShut {NoStop}%
\bibitem [{\citenamefont {Zou}\ \emph {et~al.}(2014)\citenamefont {Zou},
  \citenamefont {Liu}, \citenamefont {Lai}, \citenamefont {Unmuth-Yockey},
  \citenamefont {Yang}, \citenamefont {Bazavov}, \citenamefont {Xie},
  \citenamefont {Xiang}, \citenamefont {Chandrasekharan}, \citenamefont
  {Tsai},\ and\ \citenamefont {Meurice}}]{PhysRevA.90.063603}%
  \BibitemOpen
  \bibfield  {author} {\bibinfo {author} {\bibfnamefont {H.}~\bibnamefont
  {Zou}}, \bibinfo {author} {\bibfnamefont {Y.}~\bibnamefont {Liu}}, \bibinfo
  {author} {\bibfnamefont {C.-Y.}\ \bibnamefont {Lai}}, \bibinfo {author}
  {\bibfnamefont {J.}~\bibnamefont {Unmuth-Yockey}}, \bibinfo {author}
  {\bibfnamefont {L.-P.}\ \bibnamefont {Yang}}, \bibinfo {author}
  {\bibfnamefont {A.}~\bibnamefont {Bazavov}}, \bibinfo {author} {\bibfnamefont
  {Z.~Y.}\ \bibnamefont {Xie}}, \bibinfo {author} {\bibfnamefont
  {T.}~\bibnamefont {Xiang}}, \bibinfo {author} {\bibfnamefont
  {S.}~\bibnamefont {Chandrasekharan}}, \bibinfo {author} {\bibfnamefont
  {S.-W.}\ \bibnamefont {Tsai}}, \ and\ \bibinfo {author} {\bibfnamefont
  {Y.}~\bibnamefont {Meurice}},\ }\href {\doibase 10.1103/PhysRevA.90.063603}
  {\bibfield  {journal} {\bibinfo  {journal} {Phys. Rev. A}\ }\textbf {\bibinfo
  {volume} {90}},\ \bibinfo {pages} {063603} (\bibinfo {year}
  {2014})}\BibitemShut {NoStop}%
\bibitem [{\citenamefont {Bazavov}\ \emph {et~al.}(2015)\citenamefont
  {Bazavov}, \citenamefont {Meurice}, \citenamefont {Tsai}, \citenamefont
  {Unmuth-Yockey},\ and\ \citenamefont {Zhang}}]{PhysRevD.92.076003}%
  \BibitemOpen
  \bibfield  {author} {\bibinfo {author} {\bibfnamefont {A.}~\bibnamefont
  {Bazavov}}, \bibinfo {author} {\bibfnamefont {Y.}~\bibnamefont {Meurice}},
  \bibinfo {author} {\bibfnamefont {S.-W.}\ \bibnamefont {Tsai}}, \bibinfo
  {author} {\bibfnamefont {J.}~\bibnamefont {Unmuth-Yockey}}, \ and\ \bibinfo
  {author} {\bibfnamefont {J.}~\bibnamefont {Zhang}},\ }\href {\doibase
  10.1103/PhysRevD.92.076003} {\bibfield  {journal} {\bibinfo  {journal} {Phys.
  Rev. D}\ }\textbf {\bibinfo {volume} {92}},\ \bibinfo {pages} {076003}
  (\bibinfo {year} {2015})}\BibitemShut {NoStop}%
\bibitem [{\citenamefont {Unmuth-Yockey}\ \emph {et~al.}(2018)\citenamefont
  {Unmuth-Yockey}, \citenamefont {Zhang}, \citenamefont {Bazavov},
  \citenamefont {Meurice},\ and\ \citenamefont {Tsai}}]{PhysRevD.98.094511}%
  \BibitemOpen
  \bibfield  {author} {\bibinfo {author} {\bibfnamefont {J.}~\bibnamefont
  {Unmuth-Yockey}}, \bibinfo {author} {\bibfnamefont {J.}~\bibnamefont
  {Zhang}}, \bibinfo {author} {\bibfnamefont {A.}~\bibnamefont {Bazavov}},
  \bibinfo {author} {\bibfnamefont {Y.}~\bibnamefont {Meurice}}, \ and\
  \bibinfo {author} {\bibfnamefont {S.-W.}\ \bibnamefont {Tsai}},\ }\href
  {\doibase 10.1103/PhysRevD.98.094511} {\bibfield  {journal} {\bibinfo
  {journal} {Phys. Rev. D}\ }\textbf {\bibinfo {volume} {98}},\ \bibinfo
  {pages} {094511} (\bibinfo {year} {2018})}\BibitemShut {NoStop}%
\bibitem [{\citenamefont {Campos~Venuti}\ and\ \citenamefont
  {Zanardi}(2007)}]{PhysRevLett.99.095701}%
  \BibitemOpen
  \bibfield  {author} {\bibinfo {author} {\bibfnamefont {L.}~\bibnamefont
  {Campos~Venuti}}\ and\ \bibinfo {author} {\bibfnamefont {P.}~\bibnamefont
  {Zanardi}},\ }\href {\doibase 10.1103/PhysRevLett.99.095701} {\bibfield
  {journal} {\bibinfo  {journal} {Phys. Rev. Lett.}\ }\textbf {\bibinfo
  {volume} {99}},\ \bibinfo {pages} {095701} (\bibinfo {year}
  {2007})}\BibitemShut {NoStop}%
\bibitem [{\citenamefont {GU}(2010)}]{doi:10.1142/S0217979210056335}%
  \BibitemOpen
  \bibfield  {author} {\bibinfo {author} {\bibfnamefont {S.-J.}\ \bibnamefont
  {GU}},\ }\href {\doibase 10.1142/S0217979210056335} {\bibfield  {journal}
  {\bibinfo  {journal} {International Journal of Modern Physics B}\ }\textbf
  {\bibinfo {volume} {24}},\ \bibinfo {pages} {4371} (\bibinfo {year}
  {2010})},\ \Eprint
  {http://arxiv.org/abs/https://doi.org/10.1142/S0217979210056335}
  {https://doi.org/10.1142/S0217979210056335} \BibitemShut {NoStop}%
\bibitem [{\citenamefont {You}\ and\ \citenamefont {He}(2015)}]{You_2015}%
  \BibitemOpen
  \bibfield  {author} {\bibinfo {author} {\bibfnamefont {W.-L.}\ \bibnamefont
  {You}}\ and\ \bibinfo {author} {\bibfnamefont {L.}~\bibnamefont {He}},\
  }\href {\doibase 10.1088/0953-8984/27/20/205601} {\bibfield  {journal}
  {\bibinfo  {journal} {Journal of Physics: Condensed Matter}\ }\textbf
  {\bibinfo {volume} {27}},\ \bibinfo {pages} {205601} (\bibinfo {year}
  {2015})}\BibitemShut {NoStop}%
\bibitem [{\citenamefont {Cincio}\ \emph {et~al.}(2019)\citenamefont {Cincio},
  \citenamefont {Rams}, \citenamefont {Dziarmaga},\ and\ \citenamefont
  {Zurek}}]{PhysRevB.100.081108}%
  \BibitemOpen
  \bibfield  {author} {\bibinfo {author} {\bibfnamefont {L.}~\bibnamefont
  {Cincio}}, \bibinfo {author} {\bibfnamefont {M.~M.}\ \bibnamefont {Rams}},
  \bibinfo {author} {\bibfnamefont {J.}~\bibnamefont {Dziarmaga}}, \ and\
  \bibinfo {author} {\bibfnamefont {W.~H.}\ \bibnamefont {Zurek}},\ }\href
  {\doibase 10.1103/PhysRevB.100.081108} {\bibfield  {journal} {\bibinfo
  {journal} {Phys. Rev. B}\ }\textbf {\bibinfo {volume} {100}},\ \bibinfo
  {pages} {081108} (\bibinfo {year} {2019})}\BibitemShut {NoStop}%
\bibitem [{\citenamefont {Eisert}\ \emph {et~al.}(2010)\citenamefont {Eisert},
  \citenamefont {Cramer},\ and\ \citenamefont {Plenio}}]{RevModPhys.82.277}%
  \BibitemOpen
  \bibfield  {author} {\bibinfo {author} {\bibfnamefont {J.}~\bibnamefont
  {Eisert}}, \bibinfo {author} {\bibfnamefont {M.}~\bibnamefont {Cramer}}, \
  and\ \bibinfo {author} {\bibfnamefont {M.~B.}\ \bibnamefont {Plenio}},\
  }\href {\doibase 10.1103/RevModPhys.82.277} {\bibfield  {journal} {\bibinfo
  {journal} {Rev. Mod. Phys.}\ }\textbf {\bibinfo {volume} {82}},\ \bibinfo
  {pages} {277} (\bibinfo {year} {2010})}\BibitemShut {NoStop}%
\bibitem [{\citenamefont {Holzhey}\ \emph {et~al.}(1994)\citenamefont
  {Holzhey}, \citenamefont {Larsen},\ and\ \citenamefont
  {Wilczek}}]{HOLZHEY1994443}%
  \BibitemOpen
  \bibfield  {author} {\bibinfo {author} {\bibfnamefont {C.}~\bibnamefont
  {Holzhey}}, \bibinfo {author} {\bibfnamefont {F.}~\bibnamefont {Larsen}}, \
  and\ \bibinfo {author} {\bibfnamefont {F.}~\bibnamefont {Wilczek}},\ }\href
  {\doibase https://doi.org/10.1016/0550-3213(94)90402-2} {\bibfield  {journal}
  {\bibinfo  {journal} {Nuclear Physics B}\ }\textbf {\bibinfo {volume}
  {424}},\ \bibinfo {pages} {443} (\bibinfo {year} {1994})}\BibitemShut
  {NoStop}%
\bibitem [{\citenamefont {Vidal}\ \emph {et~al.}(2003)\citenamefont {Vidal},
  \citenamefont {Latorre}, \citenamefont {Rico},\ and\ \citenamefont
  {Kitaev}}]{PhysRevLett.90.227902}%
  \BibitemOpen
  \bibfield  {author} {\bibinfo {author} {\bibfnamefont {G.}~\bibnamefont
  {Vidal}}, \bibinfo {author} {\bibfnamefont {J.~I.}\ \bibnamefont {Latorre}},
  \bibinfo {author} {\bibfnamefont {E.}~\bibnamefont {Rico}}, \ and\ \bibinfo
  {author} {\bibfnamefont {A.}~\bibnamefont {Kitaev}},\ }\href {\doibase
  10.1103/PhysRevLett.90.227902} {\bibfield  {journal} {\bibinfo  {journal}
  {Phys. Rev. Lett.}\ }\textbf {\bibinfo {volume} {90}},\ \bibinfo {pages}
  {227902} (\bibinfo {year} {2003})}\BibitemShut {NoStop}%
\bibitem [{\citenamefont {White}(1992)}]{PhysRevLett.69.2863}%
  \BibitemOpen
  \bibfield  {author} {\bibinfo {author} {\bibfnamefont {S.~R.}\ \bibnamefont
  {White}},\ }\href {\doibase 10.1103/PhysRevLett.69.2863} {\bibfield
  {journal} {\bibinfo  {journal} {Phys. Rev. Lett.}\ }\textbf {\bibinfo
  {volume} {69}},\ \bibinfo {pages} {2863} (\bibinfo {year}
  {1992})}\BibitemShut {NoStop}%
\bibitem [{\citenamefont {White}(1993)}]{PhysRevB.48.10345}%
  \BibitemOpen
  \bibfield  {author} {\bibinfo {author} {\bibfnamefont {S.~R.}\ \bibnamefont
  {White}},\ }\href {\doibase 10.1103/PhysRevB.48.10345} {\bibfield  {journal}
  {\bibinfo  {journal} {Phys. Rev. B}\ }\textbf {\bibinfo {volume} {48}},\
  \bibinfo {pages} {10345} (\bibinfo {year} {1993})}\BibitemShut {NoStop}%
\bibitem [{\citenamefont {SchollwÃ¶ck}(2011)}]{SCHOLLWOCK201196}%
  \BibitemOpen
  \bibfield  {author} {\bibinfo {author} {\bibfnamefont {U.}~\bibnamefont
  {SchollwÃ¶ck}},\ }\href {\doibase
  https://doi.org/10.1016/j.aop.2010.09.012} {\bibfield  {journal} {\bibinfo
  {journal} {Annals of Physics}\ }\textbf {\bibinfo {volume} {326}},\ \bibinfo
  {pages} {96 } (\bibinfo {year} {2011})},\ \bibinfo {note} {january 2011
  Special Issue}\BibitemShut {NoStop}%
\bibitem [{\citenamefont {Fishman}\ \emph {et~al.}(2020)\citenamefont
  {Fishman}, \citenamefont {White},\ and\ \citenamefont
  {Stoudenmire}}]{itensor}%
  \BibitemOpen
  \bibfield  {author} {\bibinfo {author} {\bibfnamefont {M.}~\bibnamefont
  {Fishman}}, \bibinfo {author} {\bibfnamefont {S.~R.}\ \bibnamefont {White}},
  \ and\ \bibinfo {author} {\bibfnamefont {E.~M.}\ \bibnamefont
  {Stoudenmire}},\ }\href@noop {} {\enquote {\bibinfo {title} {The
  \mbox{ITensor} software library for tensor network calculations},}\ }
  (\bibinfo {year} {2020}),\ \Eprint {http://arxiv.org/abs/2007.14822}
  {arXiv:2007.14822} \BibitemShut {NoStop}%
\bibitem [{\citenamefont {\"Ostlund}\ and\ \citenamefont
  {Rommer}(1995)}]{PhysRevLett.75.3537}%
  \BibitemOpen
  \bibfield  {author} {\bibinfo {author} {\bibfnamefont {S.}~\bibnamefont
  {\"Ostlund}}\ and\ \bibinfo {author} {\bibfnamefont {S.}~\bibnamefont
  {Rommer}},\ }\href {\doibase 10.1103/PhysRevLett.75.3537} {\bibfield
  {journal} {\bibinfo  {journal} {Phys. Rev. Lett.}\ }\textbf {\bibinfo
  {volume} {75}},\ \bibinfo {pages} {3537} (\bibinfo {year}
  {1995})}\BibitemShut {NoStop}%
\bibitem [{\citenamefont {White}(2005)}]{PhysRevB.72.180403}%
  \BibitemOpen
  \bibfield  {author} {\bibinfo {author} {\bibfnamefont {S.~R.}\ \bibnamefont
  {White}},\ }\href {\doibase 10.1103/PhysRevB.72.180403} {\bibfield  {journal}
  {\bibinfo  {journal} {Phys. Rev. B}\ }\textbf {\bibinfo {volume} {72}},\
  \bibinfo {pages} {180403} (\bibinfo {year} {2005})}\BibitemShut {NoStop}%
\bibitem [{\citenamefont {{Gu, Shi-Jian}}\ and\ \citenamefont {{Yu, Wing
  Chi}}(2014)}]{SJGu2014}%
  \BibitemOpen
  \bibfield  {author} {\bibinfo {author} {\bibnamefont {{Gu, Shi-Jian}}}\ and\
  \bibinfo {author} {\bibnamefont {{Yu, Wing Chi}}},\ }\href {\doibase
  10.1209/0295-5075/108/20002} {\bibfield  {journal} {\bibinfo  {journal}
  {EPL}\ }\textbf {\bibinfo {volume} {108}},\ \bibinfo {pages} {20002}
  (\bibinfo {year} {2014})}\BibitemShut {NoStop}%
\bibitem [{\citenamefont {Hsieh}\ \emph {et~al.}(2013)\citenamefont {Hsieh},
  \citenamefont {Kao},\ and\ \citenamefont {Sandvik}}]{Hsieh_2013}%
  \BibitemOpen
  \bibfield  {author} {\bibinfo {author} {\bibfnamefont {Y.-D.}\ \bibnamefont
  {Hsieh}}, \bibinfo {author} {\bibfnamefont {Y.-J.}\ \bibnamefont {Kao}}, \
  and\ \bibinfo {author} {\bibfnamefont {A.~W.}\ \bibnamefont {Sandvik}},\
  }\href {\doibase 10.1088/1742-5468/2013/09/p09001} {\bibfield  {journal}
  {\bibinfo  {journal} {Journal of Statistical Mechanics: Theory and
  Experiment}\ }\textbf {\bibinfo {volume} {2013}},\ \bibinfo {pages} {P09001}
  (\bibinfo {year} {2013})}\BibitemShut {NoStop}%
\bibitem [{\citenamefont {Wallin}\ and\ \citenamefont
  {Weber}(1995)}]{PhysRevB.51.6163}%
  \BibitemOpen
  \bibfield  {author} {\bibinfo {author} {\bibfnamefont {M.}~\bibnamefont
  {Wallin}}\ and\ \bibinfo {author} {\bibfnamefont {H.}~\bibnamefont {Weber}},\
  }\href {\doibase 10.1103/PhysRevB.51.6163} {\bibfield  {journal} {\bibinfo
  {journal} {Phys. Rev. B}\ }\textbf {\bibinfo {volume} {51}},\ \bibinfo
  {pages} {6163} (\bibinfo {year} {1995})}\BibitemShut {NoStop}%
\bibitem [{\citenamefont {Carrasquilla}\ \emph {et~al.}(2013)\citenamefont
  {Carrasquilla}, \citenamefont {Manmana},\ and\ \citenamefont
  {Rigol}}]{PhysRevA.87.043606}%
  \BibitemOpen
  \bibfield  {author} {\bibinfo {author} {\bibfnamefont {J.}~\bibnamefont
  {Carrasquilla}}, \bibinfo {author} {\bibfnamefont {S.~R.}\ \bibnamefont
  {Manmana}}, \ and\ \bibinfo {author} {\bibfnamefont {M.}~\bibnamefont
  {Rigol}},\ }\href {\doibase 10.1103/PhysRevA.87.043606} {\bibfield  {journal}
  {\bibinfo  {journal} {Phys. Rev. A}\ }\textbf {\bibinfo {volume} {87}},\
  \bibinfo {pages} {043606} (\bibinfo {year} {2013})}\BibitemShut {NoStop}%
\bibitem [{\citenamefont {Dalmonte}\ \emph {et~al.}(2015)\citenamefont
  {Dalmonte}, \citenamefont {Carrasquilla}, \citenamefont {Taddia},
  \citenamefont {Ercolessi},\ and\ \citenamefont {Rigol}}]{PhysRevB.91.165136}%
  \BibitemOpen
  \bibfield  {author} {\bibinfo {author} {\bibfnamefont {M.}~\bibnamefont
  {Dalmonte}}, \bibinfo {author} {\bibfnamefont {J.}~\bibnamefont
  {Carrasquilla}}, \bibinfo {author} {\bibfnamefont {L.}~\bibnamefont
  {Taddia}}, \bibinfo {author} {\bibfnamefont {E.}~\bibnamefont {Ercolessi}}, \
  and\ \bibinfo {author} {\bibfnamefont {M.}~\bibnamefont {Rigol}},\ }\href
  {\doibase 10.1103/PhysRevB.91.165136} {\bibfield  {journal} {\bibinfo
  {journal} {Phys. Rev. B}\ }\textbf {\bibinfo {volume} {91}},\ \bibinfo
  {pages} {165136} (\bibinfo {year} {2015})}\BibitemShut {NoStop}%
\bibitem [{\citenamefont {Nomura}\ and\ \citenamefont
  {Kitazawa}(1998)}]{Nomura_1998}%
  \BibitemOpen
  \bibfield  {author} {\bibinfo {author} {\bibfnamefont {K.}~\bibnamefont
  {Nomura}}\ and\ \bibinfo {author} {\bibfnamefont {A.}~\bibnamefont
  {Kitazawa}},\ }\href {\doibase 10.1088/0305-4470/31/36/008} {\bibfield
  {journal} {\bibinfo  {journal} {Journal of Physics A: Mathematical and
  General}\ }\textbf {\bibinfo {volume} {31}},\ \bibinfo {pages} {7341}
  (\bibinfo {year} {1998})}\BibitemShut {NoStop}%
\bibitem [{\citenamefont {Manmana}\ \emph {et~al.}(2011)\citenamefont
  {Manmana}, \citenamefont {Hazzard}, \citenamefont {Chen}, \citenamefont
  {Feiguin},\ and\ \citenamefont {Rey}}]{PhysRevA.84.043601}%
  \BibitemOpen
  \bibfield  {author} {\bibinfo {author} {\bibfnamefont {S.~R.}\ \bibnamefont
  {Manmana}}, \bibinfo {author} {\bibfnamefont {K.~R.~A.}\ \bibnamefont
  {Hazzard}}, \bibinfo {author} {\bibfnamefont {G.}~\bibnamefont {Chen}},
  \bibinfo {author} {\bibfnamefont {A.~E.}\ \bibnamefont {Feiguin}}, \ and\
  \bibinfo {author} {\bibfnamefont {A.~M.}\ \bibnamefont {Rey}},\ }\href
  {\doibase 10.1103/PhysRevA.84.043601} {\bibfield  {journal} {\bibinfo
  {journal} {Phys. Rev. A}\ }\textbf {\bibinfo {volume} {84}},\ \bibinfo
  {pages} {043601} (\bibinfo {year} {2011})}\BibitemShut {NoStop}%
\bibitem [{\citenamefont {Wang}\ \emph {et~al.}(2010)\citenamefont {Wang},
  \citenamefont {Feng},\ and\ \citenamefont {Chen}}]{PhysRevA.81.064301}%
  \BibitemOpen
  \bibfield  {author} {\bibinfo {author} {\bibfnamefont {B.}~\bibnamefont
  {Wang}}, \bibinfo {author} {\bibfnamefont {M.}~\bibnamefont {Feng}}, \ and\
  \bibinfo {author} {\bibfnamefont {Z.-Q.}\ \bibnamefont {Chen}},\ }\href
  {\doibase 10.1103/PhysRevA.81.064301} {\bibfield  {journal} {\bibinfo
  {journal} {Phys. Rev. A}\ }\textbf {\bibinfo {volume} {81}},\ \bibinfo
  {pages} {064301} (\bibinfo {year} {2010})}\BibitemShut {NoStop}%
\bibitem [{\citenamefont {Rigol}\ \emph {et~al.}(2009)\citenamefont {Rigol},
  \citenamefont {Shastry},\ and\ \citenamefont {Haas}}]{PhysRevB.80.094529}%
  \BibitemOpen
  \bibfield  {author} {\bibinfo {author} {\bibfnamefont {M.}~\bibnamefont
  {Rigol}}, \bibinfo {author} {\bibfnamefont {B.~S.}\ \bibnamefont {Shastry}},
  \ and\ \bibinfo {author} {\bibfnamefont {S.}~\bibnamefont {Haas}},\ }\href
  {\doibase 10.1103/PhysRevB.80.094529} {\bibfield  {journal} {\bibinfo
  {journal} {Phys. Rev. B}\ }\textbf {\bibinfo {volume} {80}},\ \bibinfo
  {pages} {094529} (\bibinfo {year} {2009})}\BibitemShut {NoStop}%
\bibitem [{\citenamefont {You}\ and\ \citenamefont
  {Dong}(2011)}]{PhysRevB.84.174426}%
  \BibitemOpen
  \bibfield  {author} {\bibinfo {author} {\bibfnamefont {W.-L.}\ \bibnamefont
  {You}}\ and\ \bibinfo {author} {\bibfnamefont {Y.-L.}\ \bibnamefont {Dong}},\
  }\href {\doibase 10.1103/PhysRevB.84.174426} {\bibfield  {journal} {\bibinfo
  {journal} {Phys. Rev. B}\ }\textbf {\bibinfo {volume} {84}},\ \bibinfo
  {pages} {174426} (\bibinfo {year} {2011})}\BibitemShut {NoStop}%
\bibitem [{\citenamefont {Jia}\ \emph {et~al.}(2011)\citenamefont {Jia},
  \citenamefont {Moritz}, \citenamefont {Chen}, \citenamefont {Shastry},\ and\
  \citenamefont {Devereaux}}]{PhysRevB.84.125113}%
  \BibitemOpen
  \bibfield  {author} {\bibinfo {author} {\bibfnamefont {C.~J.}\ \bibnamefont
  {Jia}}, \bibinfo {author} {\bibfnamefont {B.}~\bibnamefont {Moritz}},
  \bibinfo {author} {\bibfnamefont {C.-C.}\ \bibnamefont {Chen}}, \bibinfo
  {author} {\bibfnamefont {B.~S.}\ \bibnamefont {Shastry}}, \ and\ \bibinfo
  {author} {\bibfnamefont {T.~P.}\ \bibnamefont {Devereaux}},\ }\href {\doibase
  10.1103/PhysRevB.84.125113} {\bibfield  {journal} {\bibinfo  {journal} {Phys.
  Rev. B}\ }\textbf {\bibinfo {volume} {84}},\ \bibinfo {pages} {125113}
  (\bibinfo {year} {2011})}\BibitemShut {NoStop}%
\bibitem [{\citenamefont {Luo}\ \emph {et~al.}(2014)\citenamefont {Luo},
  \citenamefont {Zhou}, \citenamefont {Liu}, \citenamefont {Liang},\ and\
  \citenamefont {Zhang}}]{PhysRevA.89.043612}%
  \BibitemOpen
  \bibfield  {author} {\bibinfo {author} {\bibfnamefont {X.}~\bibnamefont
  {Luo}}, \bibinfo {author} {\bibfnamefont {K.}~\bibnamefont {Zhou}}, \bibinfo
  {author} {\bibfnamefont {W.}~\bibnamefont {Liu}}, \bibinfo {author}
  {\bibfnamefont {Z.}~\bibnamefont {Liang}}, \ and\ \bibinfo {author}
  {\bibfnamefont {Z.}~\bibnamefont {Zhang}},\ }\href {\doibase
  10.1103/PhysRevA.89.043612} {\bibfield  {journal} {\bibinfo  {journal} {Phys.
  Rev. A}\ }\textbf {\bibinfo {volume} {89}},\ \bibinfo {pages} {043612}
  (\bibinfo {year} {2014})}\BibitemShut {NoStop}%
\bibitem [{\citenamefont {Nishiyama}(2019)}]{Nishiyama2019}%
  \BibitemOpen
  \bibfield  {author} {\bibinfo {author} {\bibfnamefont {Y.}~\bibnamefont
  {Nishiyama}},\ }\href {\doibase 10.1140/epjb/e2019-100269-8} {\bibfield
  {journal} {\bibinfo  {journal} {The European Physical Journal B}\ }\textbf
  {\bibinfo {volume} {92}},\ \bibinfo {pages} {167} (\bibinfo {year}
  {2019})}\BibitemShut {NoStop}%
\bibitem [{\citenamefont {Sirker}(2010)}]{PhysRevLett.105.117203}%
  \BibitemOpen
  \bibfield  {author} {\bibinfo {author} {\bibfnamefont {J.}~\bibnamefont
  {Sirker}},\ }\href {\doibase 10.1103/PhysRevLett.105.117203} {\bibfield
  {journal} {\bibinfo  {journal} {Phys. Rev. Lett.}\ }\textbf {\bibinfo
  {volume} {105}},\ \bibinfo {pages} {117203} (\bibinfo {year}
  {2010})}\BibitemShut {NoStop}%
\bibitem [{\citenamefont {Albuquerque}\ \emph {et~al.}(2010)\citenamefont
  {Albuquerque}, \citenamefont {Alet}, \citenamefont {Sire},\ and\
  \citenamefont {Capponi}}]{PhysRevB.81.064418}%
  \BibitemOpen
  \bibfield  {author} {\bibinfo {author} {\bibfnamefont {A.~F.}\ \bibnamefont
  {Albuquerque}}, \bibinfo {author} {\bibfnamefont {F.}~\bibnamefont {Alet}},
  \bibinfo {author} {\bibfnamefont {C.}~\bibnamefont {Sire}}, \ and\ \bibinfo
  {author} {\bibfnamefont {S.}~\bibnamefont {Capponi}},\ }\href {\doibase
  10.1103/PhysRevB.81.064418} {\bibfield  {journal} {\bibinfo  {journal} {Phys.
  Rev. B}\ }\textbf {\bibinfo {volume} {81}},\ \bibinfo {pages} {064418}
  (\bibinfo {year} {2010})}\BibitemShut {NoStop}%
\bibitem [{\citenamefont {Wang}\ \emph {et~al.}(2015)\citenamefont {Wang},
  \citenamefont {Liu}, \citenamefont {Imri\ifmmode~\check{s}\else
  \v{s}\fi{}ka}, \citenamefont {Ma},\ and\ \citenamefont
  {Troyer}}]{PhysRevX.5.031007}%
  \BibitemOpen
  \bibfield  {author} {\bibinfo {author} {\bibfnamefont {L.}~\bibnamefont
  {Wang}}, \bibinfo {author} {\bibfnamefont {Y.-H.}\ \bibnamefont {Liu}},
  \bibinfo {author} {\bibfnamefont {J.}~\bibnamefont
  {Imri\ifmmode~\check{s}\else \v{s}\fi{}ka}}, \bibinfo {author} {\bibfnamefont
  {P.~N.}\ \bibnamefont {Ma}}, \ and\ \bibinfo {author} {\bibfnamefont
  {M.}~\bibnamefont {Troyer}},\ }\href {\doibase 10.1103/PhysRevX.5.031007}
  {\bibfield  {journal} {\bibinfo  {journal} {Phys. Rev. X}\ }\textbf {\bibinfo
  {volume} {5}},\ \bibinfo {pages} {031007} (\bibinfo {year}
  {2015})}\BibitemShut {NoStop}%
\end{thebibliography}
%

\end{document}